\documentclass[twoside,a4paper]{IEEEtran}

\usepackage[noadjust]{cite} % Improves presentation of citations, like [3--7].
\usepackage{multirow}
\usepackage{amsmath}
\usepackage{overpic}
\usepackage{amsopn}
\usepackage{amssymb}
\usepackage{lettrine}
\usepackage{accents}
\usepackage{mathrsfs}
\usepackage{pifont}

\usepackage{bm}         % for nice bold math symbols using \bm{...}
\usepackage{fix-cm}     % permit arbitrary sizes of Computer Modern
\usepackage{algorithm}
\usepackage{algorithmicx}
\usepackage{algpseudocode}

\makeatletter
\newcommand{\removelatexerror}{\let\@latex@error\@gobble}
\makeatother

\usepackage{mleftright} %provides \mleft, \mright to fix issue with spacing
\mleftright                 %redefines all \left and \right to behave
\usepackage{mathdots}   % fixes \ddots and \vdots to scale properly

\makeatletter
\def\IEEElabelanchoreqn#1{\bgroup
\def\@currentlabel{\p@equation\theequation}\relax
\def\@currentHref{\@IEEEtheHrefequation}\label{#1}\relax
\Hy@raisedlink{\hyper@anchorstart{\@currentHref}}\relax
\Hy@raisedlink{\hyper@anchorend}\egroup}
\makeatother

\usepackage[utf8]{inputenc} 
\usepackage[T1]{fontenc} % To force 8-bit encoding so that
\usepackage{tikz}
\usepgflibrary{arrows.meta}
\usetikzlibrary{positioning}
\usetikzlibrary{shadows} %used also in mdframed
\usetikzlibrary{calc}
\usetikzlibrary{patterns}
\usepackage{pgfplots}
\pgfplotsset{compat=1.15}
\usepgfplotslibrary{units}
\usetikzlibrary{spy,backgrounds}
\usetikzlibrary{decorations.markings}
\usepackage{pgfplotstable}

\newtheorem{theorem}{Theorem}
\newtheorem{lemma}{Lemma}
\newtheorem{corollary}{Corollary}

\newtheorem{proposition}{Proposition}

\newtheorem{remark}{Remark}
\newtheorem{definition}{Definition}

\newcommand{\snr}{\mathsf{snr}} 

\newcommand{\remarksymbol}{\text{$\blacktriangle$}}

\usepackage{hyphenat} % Package for hypenation-stuff

\newcommand{\expec}[1]{\mathbb{E}{\left[#1\right]}}

\newcommand{\midk}[1]{\kern0.1em #1 \kern0.1em}
\newcommand{\middlek}[1]{\kern0.1em \middle#1 \kern0.1em}
\newcommand{\bigk}[1]{\kern-0.1em \bigm#1 \kern-0.1em}
\newcommand{\Bigk}[1]{\kern-0.1em \Bigm#1 \kern-0.1em}
\newcommand{\biggk}[1]{\kern-0.1em \biggm#1 \kern-0.1em}
\newcommand{\Biggk}[1]{\kern-0.1em \Biggm#1 \kern-0.1em}

\newcommand{\tn}[1]{\textnormal{#1}}

\newcommand{\II}{\mathop{}\!\const{I}}
\newcommand{\mat}[1]{\mathbb{#1}}
\newcommand{\trans}[1]{#1^{\textnormal{\textsf{\tiny T}}}} % transpose  
\newcommand{\hermi}[1]{#1^{\dagger}} % transpose and conjugate
\newcommand{\trace}[1]{\operatorname{tr}\left(#1\right)} % trace

\newcommand{\rp}[1]{\mathfrak{Re}\left\{#1\right\}}  
\newcommand{\ip}[1]{\mathfrak{Im}\left\{#1\right\}}  
\newcommand{\QED}{\hfill$\blacksquare$}

\newcommand{\rev}[1]{{\color{black}#1}}

\newcommand{\const}[1]{\textnormal{\usefont{U}{eur}{m}{n}\selectfont #1}} % Euler
\newcommand{\hh}{\mathop{}\!\const{h}}  % differential entropy
\newcommand{\relD}{\mathop{}\!\mathsf{D}}         % relative entropy
\newcommand{\relDf}[2]{\relD\left(#1 \kern0.1em\middle\|\kern0.1em #2\right)}
\newcommand{\erelDf}[2]{\relD(#1 \kern0.1em\|\kern0.1em #2)} 
\newcommand{\bigrelDf}[2]{\relD\bigl(#1 \kern-0.1em \bigm\| \kern-0.1em#2\bigr)}
\newcommand{\BigrelDf}[2]{\relD\Bigl(#1 \kern-0.1em \Bigm\| \kern-0.1em#2\Bigr)}
\newcommand{\biggrelDf}[2]{\relD\biggl(#1 \kern-0.1em \biggm\| \kern-0.1em#2\biggr)}
\newcommand{\BiggrelDf}[2]{\relD\Biggl(#1 \kern-0.1em \Biggm\| \kern-0.1em#2\Biggr)}

\newcommand{\nt}{n_{\tn{T}}}
\newcommand{\nr}{n_{\tn{R}}}

\newcommand{\Prvcond}[2]{\Pr\left[#1 \kern0.1em\middle|\kern0.1em #2\right]}
\newcommand{\ePrvcond}[2]{\Pr[#1 \kern0.1em|\kern0.1em #2]} 
\newcommand{\bigPrvcond}[2]{\Pr\bigl[#1 \kern-0.1em \bigm| \kern-0.1em#2\bigr]}
\newcommand{\BigPrvcond}[2]{\Pr\Bigl[#1 \kern-0.1em \Bigm| \kern-0.1em#2\Bigr]}
\newcommand{\biggPrvcond}[2]{\Pr\biggl[#1 \kern-0.1em \biggm| \kern-0.1em#2\biggr]}
\newcommand{\BiggPrvcond}[2]{\Pr\Biggl[#1 \kern-0.1em \Biggm| \kern-0.1em#2\Biggr]}

\newcommand{\Prscond}[2]{\Pr\left(#1 \kern0.1em\middle|\kern0.1em #2\right)}
\newcommand{\ePrscond}[2]{\Pr(#1 \kern0.1em|\kern0.1em #2)} 
\newcommand{\bigPrscond}[2]{\Pr\bigl(#1 \kern-0.1em \bigm| \kern-0.1em#2\bigr)}
\newcommand{\BigPrscond}[2]{\Pr\Bigl(#1 \kern-0.1em \Bigm| \kern-0.1em#2\Bigr)}
\newcommand{\biggPrscond}[2]{\Pr\biggl(#1 \kern-0.1em \biggm| \kern-0.1em#2\biggr)}
\newcommand{\BiggPrscond}[2]{\Pr\Biggl(#1 \kern-0.1em \Biggm| \kern-0.1em#2\Biggr)}

\newcommand{\Exp}{\operatorname{\textnormal{\textsf{E}}}}

\newcommand{\Econd}[3][]{\Exp_{#1}\left[#2 \kern0.1em\middle|\kern0.1em #3\right]}
\newcommand{\eEcond}[3][]{\Exp_{#1}[#2 \kern0.1em|\kern0.1em #3]}
\newcommand{\bigEcond}[3][]{\Exp_{#1}\bigl[#2 \kern-0.1em \bigm| \kern-0.1em #3\bigr]}
\newcommand{\BigEcond}[3][]{\Exp_{#1}\Bigl[#2 \kern-0.1em \Bigm| \kern-0.1em #3\Bigr]}
\newcommand{\biggEcond}[3][]{\Exp_{#1}\biggl[#2 \kern-0.1em \biggm| \kern-0.1em #3\biggr]}
\newcommand{\BiggEcond}[3][]{\Exp_{#1}\Biggl[#2 \kern-0.1em \Biggm| \kern-0.1em #3\Biggr]}

\newcommand{\cov}[1]{\operatorname{\mat{K}}_{\mspace{-1.0mu}#1\mspace{-2.0mu}#1}} % covariance matrix

\newcommand{\dd}{\mathop{}\!\mathrm{d}}

\newcommand{\supp}{\operatorname{supp}}

\newcommand{\argmax}{\operatorname*{argmax}}

\makeatletter
\newcommand{\rmnum}[1]{{\romannumeral #1}}
\newcommand{\Rmnum}[1]{\expandafter\@slowromancap\romannumeral #1@}
\makeatother

\usepackage{threeparttable}
\usepackage{diagbox}
\usepackage{array}
\usepackage{boldline}
\usepackage{caption}
\usepackage{dsfont} % less troublesome bbm fonts

\usepackage{algpseudocode}

\allowdisplaybreaks
\usetikzlibrary{datavisualization} 

\usetikzlibrary{spy,backgrounds}
\usetikzlibrary{decorations.markings}
\begin{document}

\title{On RIS-Aided SIMO Gaussian Channels: Towards A Single-RF MIMO Transceiver Architecture}

\author{Ru-Han Chen, Jing Zhou, Yonggang Zhu, Kai Zhang$^\ast$
	%	,	\\
	%  Wenyi Zhang, \IEEEmembership{Senior Member,~IEEE}, and Jing Zhou, \IEEEmembership{Member,~IEEE}%
	\thanks{This work is supported in part by the National Natural Science Foundation of China under Grant No. 62131005. \textit{(Corresponding author: Kai Zhang.)}}
	\thanks{Ru-Han Chen, Yonggang Zhu, Kai Zhang are with Sixty-third Research Institute, National University of Defense Technology, Nanjing, China (e-mail: tx\_rhc22@nudt.edu.cn, zhumaka1982@163.com, zhangkai08@nudt.edu.cn).}
	\thanks{ Jing Zhou is with Dept. Computer Science and Engineering, Shaoxing University, Shaoxing, China (e-mail:jzhou@usx.edu.cn).} 
}

\maketitle

\begin{abstract}
In this paper, for a single-input multiple-output (SIMO) system aided by a passive reconfigurable intelligent surface (RIS), the joint transmission accomplished by the single transmit antenna and the RIS with multiple controllable reflective elements is considered. 
%Via
%Three operation modes, including \textit{conventional beamforming}, \textit{input-dependent beamforming}, and \textit{joint transmission}, are considered.
Relying on a general capacity upper bound derived by a maximum-trace argument, we respectively characterize the capacity of such \rev{a} channel in the low-SNR or the rank-one regimes, in which the optimal configuration of the RIS is proved to be beamforming with carefully-chosen phase shifts. To exploit the potential of modulating extra information on the RIS, based on the QR decomposition, successive interference cancellation, and a strategy named \textit{partially beamforming and partially information-carrying}, we propose a novel transceiver architecture with only a single RF front end at the transmitter, by which the considered channel can be regarded as a concatenation of a vector Gaussian channel and several phase-modulated channels. 
%To accurately evaluate achievable rates of such the transceiver with two different signaling schemes, we specially 
Especially, we investigate a class of vector Gaussian channels with a hypersphere input support constraint, and not only generalize the existing result to arbitrary-dimensional real spaces but also present its high-order capacity asymptotics, by which both  capacities of hypersphere-constrained channels and achievable rates of the proposed transceiver with two different signaling schemes can be well-approximated.
%Based on which, we characterize the asymptotic of achievable rates under two aforementioned inputs. Our proposed transceiver architecture achieves full DoF for spatially independent channels.
Information-theoretic analyses show that the transceiver architecture designed for the SIMO channel has a boosted multiplexing gain, rather than one for the conventionally-used optimized beamforming scheme.
Numerical results verify our derived asymptotics and show notable superiority of the proposed transceiver.
% over the conventional-used optimized beamforming scheme
%.
\end{abstract}


\section{Introduction}
\label{sec:introduction}
\lettrine[lines=2]{T}{argeting} ultra-high data rates, improved energy efficiency, wide-area coverage, and scalable wireless connectivity, future wireless network need to take advantage of revolutionary technologys, such as millimeter wave (mmWave) communications,  intelligent materials and novel transceiver architectures tailored for those emerging trends. Unlike existing designs that mainly focus on the sub-6 GHz frequency band, next-generation transceivers are expected to efficiently operate over broader frequency-bands. This has motivated the concept of \textit{single radio-frequency (RF) multiple-input multiple output (MIMO)}, where a single RF chain at the transmitter aided by a reconfigurable intelligent surface (RIS) is used to
% achieve an appealing trade-off between spectral efficiency and energy efficiency  and significantly 
reduce the required number of RF chains for less energy consumption
% high cost and energy consumption of high resolution analog-to-digital/digital-to-analog converters in conventional fully-digital MIMO systems
 \cite{liqiang2021MWC,Bereyhi2020icassp,Karasik2021isit,Karasik2021adaptive}.

%In conventional fully-digital multiple-input multiple-output (MIMO) systems, each antenna is connected to a dedicated radio frequency (RF) chain, 
%due to the . 
%
%
%tremendously 

%Q. Li, M. Wen and M. Di Renzo, "Single-RF MIMO: From Spatial Modulation to Metasurface-Based Modulation," in IEEE Wireless Communications, vol. 28, no. 4, pp. 88-95, August 2021, doi: 10.1109/MWC.021.2000376.

The RIS is an artificial two-dimensional metasurface consisting of numerous nearly-passive reflecting elements, each of which can impose a controllable phase shift on the incident electromagnetic wave without involving extra RF chains for signal processing \cite{chengqiang2022proceeding}.
%manipulating the electromagnetic waves
%
%is arranged mostly in sub-wavelength spacing and able to independently impose an controllable phase shift on the incident signal. 
With lower implementation cost than conventional RF components, the ability to shape wireless channel and easy fabrication with light weight and small-thickness layers, the RIS technology is considered as one of the most promising approaches for future wireless communications \cite{Renzo2020smart}. 

%Moreover,  Since the elements of an RIS reflect the incoming signals without any signal processing operations that require radio-frequency (RF) chains, 
%
%an RIS has a lower implementation cost than conventional active receivers and transmitters. 

%, and thus can be readily taken into account for joint deployment of the base states and portable RIS

%While research into RIS-aided wireless communcation such as signal processing techniques, channel modeling, antenna design, and standardization is receiving overwhelming attention, RISs are not currently well-understood from an information-theoretic perspective.  The ultimate channel capacity and available technology to achieve it is still lacked. 
%The RIS usually acts as a “focusing lens” that can be configured to reflect or refract incoming radio waves towards arbitrary angles. 
In most prior and relevant works, the RIS usually acts as a fixed and passive beamformer that optimizes phase shifts at reflective elements (i.e., reflection pattern) to be constructively superimposed upon other paths to enhance the received signal power, or destructively cancelled to mitigate co-channel interference, malicious jamming, or signal leakage to eavesdroppers, and so on \cite{zhougui2020TSP,wuqingqing2019TCOM,huangchongwen2019TWC,tangjunwen2023transmissive,tangxiao2021securing,sunyifu2022TWC}. 
%In the beamforming mode, the phase shifts at the RIS side are prescribed and fixed until the channel state changes.
 However, the effect of the reflective RIS (either passive or active) results in a (unitary) diagonal matrix multiplied in the concatenated channel model, of which the rank remains constant despite of the choose of beamforming schemes (or equivalently, constant phase patterns). Using the RIS as a beamformer does not fulfill its potential for increasing the multiplexing gain (or equivalently, the spatial
 degrees of freedom) of the communication system. 
%
%  induced channel is still an conventional SIMO AWGN channel, and hence equivalent to a scalar one with an enhanced SNR. 
%  
Based on the classic multi-antenna communication theory\footnote{For conventional multi-antenna radio-frequency communications, the mostly-used model is the linear vector channel corrupted by additive white Gaussian noise (AWGN) and with a total average-power constraint on the transmitters, which can be equivalently converted into $\tau$ (i.e., the rank of the linear channel) parallel scalar AWGN channels via singular value decomposition at the receiver side \cite{telatar99_1}.
	%\begin{flalign}
	%	&\widetilde{Y}_i=\sigma_i\widetilde{S}_i+\widetilde{Z}_i,\, i\in [r] ;\\
	%	\text{s.t. }&\sum_{i=1}^r\mathbb{E}\left[\left|\widetilde{S}_i\right|^2\right]\le \const{E},
	%\end{flalign}
%	and the well-known water-filling strategy achieves the channel capacity.
	}, despite of an optimized beamformer by judiciously adjusting phase-shifts at the RIS side, the number of parallel data streams for highly rank-deficient channels (such as mmWave channels) are still \rev{limited}.
%, especially for the single-input multiple-output (SIMO) system. 

To overcome this information-theoretic bottleneck, one can modulate the information upon the artificially manipulated wireless environment \cite{yao2023antijamming,Khandani2013media}, to which the spatiotemporally modulated RIS emerges as a powerful approach \cite{chengqiang2022proceeding}. 
%But an exact analysis of the information theoretic limits of such systems is a challenging
%task. 
%In general, the information theoretical limit of modulating through RIS has	not been fully characterized. under the restrictive assumption that 
In \cite{cheng2021dof}, the authors comprehensively investigate the multiplexing gain for almost all MIMO AWGN channels aided by an RIS in joint transmission scenarios and under the assumption that the symbol rate at the transmit  antenna and the re-configuration rate of the RIS are identical.
%the authors
%
%Assumption $L=1$. No operational transceiver framework to achieve it. The proposed method:
%
% symbol-level precoding act as constellation mapping. Numerical method is time-consuming. no tractable analysis framework. Mismatched mapping may be an obstacle to achieve full DoF. 
% The desired equivalent set of $\hat{\mathbf{Y}}$ is not given, and characterizing such the set is challenging  (the feasible region $\mathbf{Y}$ in a form of $y\bm{1}$ that the author claimed can not be used to confirm the derived Dof results.)
In \cite{Karasik2021adaptive}, achievable rates of the SIMO link jointly encoded by an RIS and a single transmitter are carried out under restrictive assumptions of blocked direct path and a finite joint input set irrelevant to the instantaneous channel vector. In \cite{yaojiacheng2023TVT,yaojiacheng2023TWC}, a novel modulation scheme that superimposing information-bearing phase offsets to the baseline phase optimized for the beamformer is proposed for the MISO and the MIMO communication systems, respectively. 
In the meantime, recent efforts also put on designing communication prototypes that experimentally verify the effectiveness of the RIS-based single-RF MIMO transceiver \cite{tangwankai2020JSAC,daijunyan2019wireless}.
%easy fabrication of the RIS, channel measurements, and  prototype implementation for microwave and mmWave bands.  
%The next-generation transceiver design should take the RIS into account both for its ability of beamfoming and information-carrying. 
%
%Despite recent endeavors in development of  . ,  a
However, some fundamental questions from the communication community are still open.  The most representative one is \textit{whether modulating extra information upon phase shifts of the RIS always bring a larger achievable rate than the beamforming scheme}.
Generally, there has been a lack of comprehensive information-theoretic comparison to figure out whether to use the RIS as an extra transmitter or a simple beamformer. Moreover, explicit achievable rates analysis as well tractable transceiver framework for the joint transmission is also required.

In this paper, we endeavor to characterize the fundamental limit to achievable rate in the scenario that a single transmit antenna and an RIS are jointly used as the signal transmitter. It is shown that insufficiency in achievable information rates of SIMO AWGN channels can be compensated by modulating extra information upon phase shifts of the RIS elements. Our main contributions include
\begin{enumerate}
	\item \textit{Capacity Results on RIS-aided SIMO Channels}: After appropriate channel reduction, we treat the original channel as a vector Gaussian one with    the equivalent input codetermined by the channel realization, phase shifts at all reflective elements, and the amplitude of the transmitted symbol at the transmit antenna. By upper-bounding the trace of the covariance matrix of \rev{the} equivalent input, we give a capacity upper bound, by which the capacity slope at low signal-to-noise-ratios (SNRs) and the exact capacity in the rank-one regime are therefore characterized. It is shown that at low-SNRs or in the rank-one case the optimal configuration of RIS is simply using the RIS as a beamformer.
	\item \textit{Transceiver Architecture for RIS-aided SIMO Channels:} For general SIMO channels, we propose a novel transceiver architecture based on QR decomposition and successive interference cancellation, where a part of reflective elements is used as a conventional beamformer, and the others as phase modulators. Non-asymptotic and asymptotic achievable rates of the proposed transceiver with two different signaling schemes are analyzed by using our improved results on the hypersphere-modulated channels. Both theoretic analysis and numerical results verifies the advantages of the proposed transceiver over the conventional beamforming-based transceiver. 
%	the insufficiency in achievable information rates of SIMO AWGN channels can be compensated by modulating extra information upon phase shifts of the RIS elements.
%	joint transmission regime Upper Bound:} 
%	\item \textit{Exact Results for Two Special Cases: Low-SNR capacity asymptotic:} The low-SNR optimal is related to a UQP problem. The optimal channel input is binary and the RIS only play the role of beamforming. 
	\item \textit{Capcity Results on Hypersphere-Modulated Channels:} We improve the existing results on the arbitrarily-dimensional vector Gaussian channel with a hypersphere input constraint, for which a modified capacity expression in a form of mathematical expectation and high-order capacity asymptotics are derived. Based on these asymptotic results, capacities of general hypersphere-modulated channels can be evaluated with good accuracy at high SNRs. 
\end{enumerate}

The rest of the paper is organized as follows. We end the introduction with notations used throughout the paper. Sec.~\ref{sec:model} introduces the model of RIS-aided SIMO AWGN channels, useful techniques for channel reduction, and the formulation of the main problem. Secs. \ref{sec:capacity_results} and \ref{sec:hp} present the derived capacity results on the RIS-aided SIMO channels and the hypersphere-modulated channels, respectively. A novel transceiver architecture and corresponding analysis of achievable rates are given in \ref{sec:qr-sic_transceiver}. Sec. \ref{sec:numerical} presents the numerical results that verify the correctness of our results. The paper is concluded and discussed in Sec.~\ref{sec:conclusion}.

\textit{Notations:} Vectors are denoted in boldface fonts, such as $x$ for a deterministic vector and $\mathbf{X}$ for a random vector. We further use another two special fonts for matrices, e.g. $\mathsf{Y}$ for deterministic matrices and $\mathbb{Y}$ for random matrices, expect for $\mathbb{E}$, $\mathbb{N}_{+}$, $\mathbb{R}$, $\mathbb{C}$ that denote the mathematical expectation operator, the sets of positive integers, real numbers, and complex numbers, respectively, as usual. The $n\times n$ identity matrix is denoted by $\mathsf{I}_n$.  For a complex-valued vector $\mathbf{x}$, $\left\| x \right\|_{p}$ denotes the $\ell_p$-norm of $\mathbf{x}$. The angle, the real part, and the imaginary part of any complex number $x\in \mathbb{C}$ are denoted by $\angle x$, $\rp{x}$ and $\ip{x}$, respectively. We use $\mathsf{vec}(\mathsf{A})$ and $\mathsf{A} \otimes \mathsf{B}$ to denote column vectorization of the matrix $\mathsf{A}$, and Kronecker product of matrices $\mathsf{A}$ and $\mathsf{B}$. For a column vector $\mathbf{x}\in \mathbb{C}^n$, $\mathsf{diag}(\mathbf{x})$ denotes a diagonal matrix with main diagonal elements $x_1$, $x_2$, $\cdots$, and $x_n$. We use $\mathcal{CN}(\cdot,\cdot)$ and $\mathcal{N}(\cdot)$ to denote the circularly symmetric complex and the real-valued Gaussian distributions, respectively. For any positive integer $n$, we define $[n]$ as the index set $\left\{1,2,\cdots,n\right\}$. $\textrm{Unif}(\mathcal{A})$ to denotes the uniform distribution on the set $\mathcal{A}$ with respect to an appropriate measure. Without specially emphasis, we let the base of logarithm be $2$.

\section{Channel Model and Problem Formulation}\label{sec:model}

\subsection{Original Model}

In this paper, we consider an RIS-aided communication link with $\nr$ receive antennas and a single transmit antenna, where an RIS with $n$ reflective element is deployed and a direct path between the transmitter and the receiver is allowed; see Fig. 1 for the system diagram.

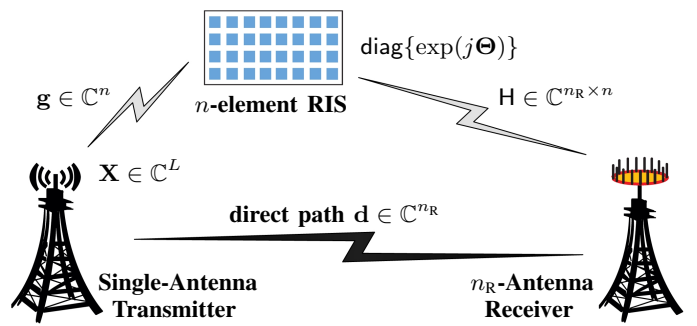
\begin{figure}[hbtp!]
	\centering
	\begin{overpic}[width=0.5\textwidth]{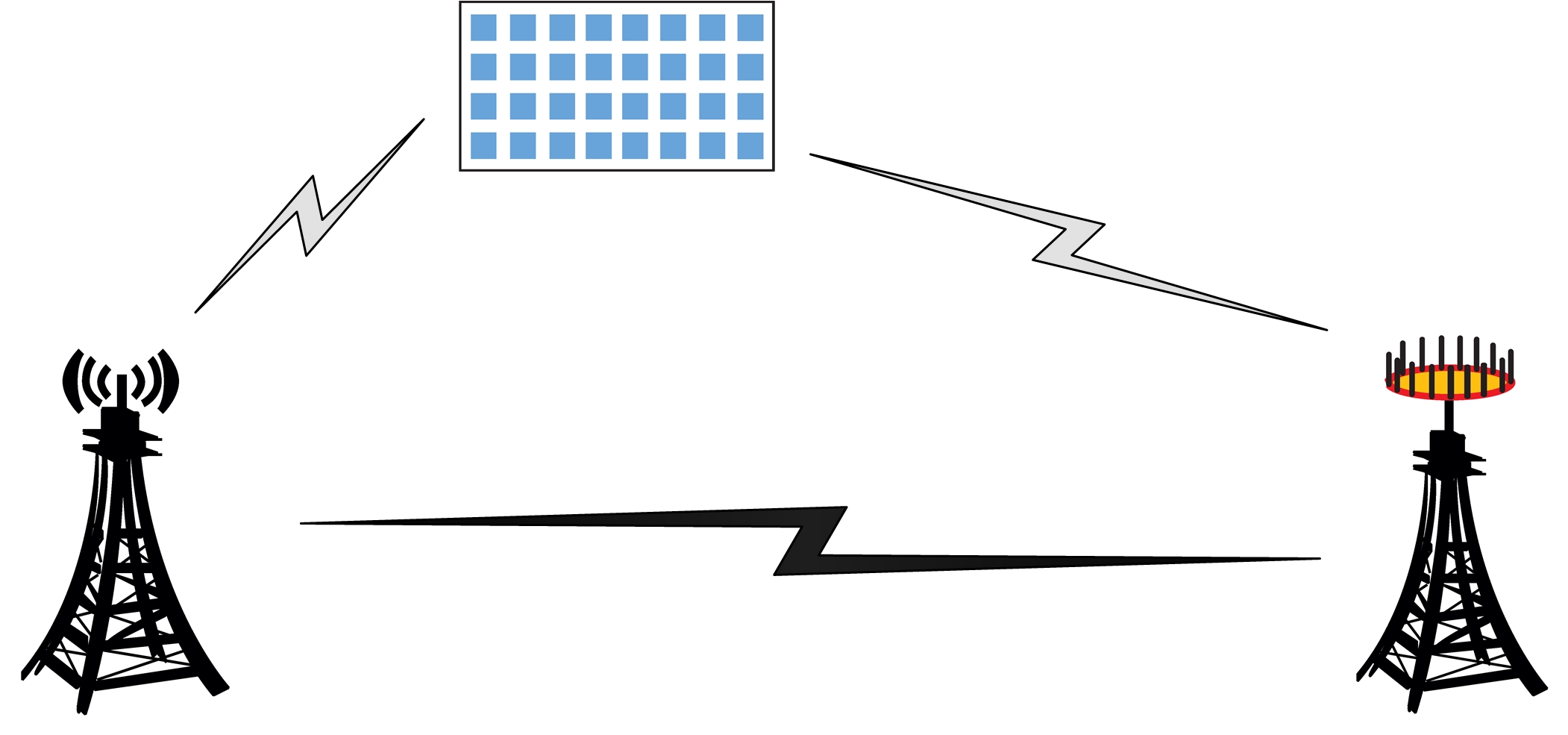}	
		\put(14,22){\small\bfseries\color{black}{$\mathbf{X}\in \mathbb{C}^L$}}
        \put(14,6){\small\bfseries\color{black}{Single-Antenna}}
        \put(16,2){\small\bfseries\color{black}{Transmitter}}
        \put(68,6){\small\bfseries\color{black}{$\nr$-Antenna}}
        \put(70,2){\small\bfseries\color{black}{Receiver}}
		\put(5,33){\small\bfseries\color{black}{$\mathbf{g}\in \mathbb{C}^{n}$}}
		\put(72,33){\small\bfseries\color{black}{$\mathsf{H}\in \mathbb{C}^{\nr\times n}$}}
		\put(33,16){\small\bfseries\color{black}{direct path $\mathbf{d}\in \mathbb{C}^{\nr}$}}
		\put(28,32){\small\bfseries\color{black}{$n$-element RIS}}
		\put(52,40.5){\small\bfseries\color{black}{$\mathsf{diag}\left\{\exp\left(j\bm{\Theta}\right)\right\}$}}
		
	\end{overpic}
	\centering \caption{Diagram of the RIS-aided SIMO communication system over a length-$L$ sub-block.}
	\label{fig:quantile}
\end{figure}

Due to practical limitation of the re-configuration rate of the RIS, we specially denote the ratio of the symbol rate at the transmitter side and the re-configuration rate of the RIS (SRR) by $L\in \mathbb{N}_{+}$. Accordingly, the phase pattern at all reflective elements remains constant in each sub-block consisting of $L$ consecutive channel uses. Then, for a static channel, its discrete-time output in a sub-block is given by
\begin{equation}\label{eqn:SIMO}
	\mathbb{Y}=
	\left(\mathsf{H}\cdot\mathsf{diag}\left\{\exp\left(j\bm{\Theta}\right)\right\}\cdot\mathbf{g} + \mathbf{d} \right)\trans{\mathbf{X}}+\mathbb{Z},
\end{equation}
 where the joint channel input $\left(\mathbf{X},\bm{\Theta} \right)$ consists of two parts:
 \begin{enumerate}
 	\item the complex-valued random column vector $\mathbf{X}\in \mathbb{C}^{L}$ denote the input signal in the sub-block from the transmit antenna, and the following average-power constraint is imposed on $\mathbf{X}$
 	\begin{flalign}\label{eq:input_constraints}
 		\frac{1}{L}\expec{\hermi{\mathbf{X}}\mathbf{X}}\le \mathcal{E},
 	\end{flalign}
 	where the positive constant $\mathcal{E}$ denotes the maximum-allowed average transmit power (per channel use);
 	\item the real-valued column vector $\bm{\Theta}=\trans{\left(\Theta_1,\cdots,\Theta_{n}\right)}\in [0,2\pi]^n$ denotes (possibly random) phase shifts at $n$ reflecting elements and the unitary diagonal matrix $\mathsf{diag}\left\{\exp\left(j\bm{\Theta}\right)\right\}$ with the main diagonal entries $\exp\left(j\Theta_{1}\right)$, $\exp\left(j\Theta_{2}\right)$, $\cdots$, and $\exp\left(j\Theta_{n}\right)$
 	denotes reflecting coefficient matrix induced by the RIS;
 \end{enumerate} 
where the matrix $\mathsf{H}\in \mathbb{C}^{\nr\times n}$, the column vector $\mathbf{g}\in \mathbb{C}^{n}$, and the column vector $\mathbf{d}\in \mathbb{C}^{\nr}$ denote deterministic channel gains from the RIS to the receive antennas, the transmitter to the RIS, and the transmitter directly to the receivers (possibly existing), respectively;  and where the $\nr\times L$ complex-valued random matrix $\mathbb{Z}=[Z_{ij}]\in \mathbb{C}^{\nr\times n}$ represents the channel noise with each component being i.i.d. standard complex Gaussian random variable, i.e., $ 	Z_{ij} \sim \mathcal{CN}(0,1)$.

\subsection{Equivalent Full-Row-Rank and Purely-Reflective Model}\label{subsec:channel_reduction}

In this subsection, an equivalent channel model is developed for simplifying subsequent analysis.

\subsubsection{Channel Merging}
According to matrix multiplication by blocks, the direct path can be merged into a new MIMO channel between the RIS and the receiver as follows
\begin{flalign}
	\mathbb{Y} &=	\left(\mathsf{H}\cdot\mathsf{diag}\left\{\exp\left(j\bm{\Theta}\right)\right\}\cdot\mathbf{g} + \mathbf{d} \right)\trans{\mathbf{X}}+\mathbb{Z}, \\
	&=\left[\mathsf{H},\mathbf{d}\right]
	\left[
	\begin{array}{cc}
		\mathsf{diag}\left\{\exp\left(j\bm{\Theta}\right)\right\}&\bm{0}_n\\
		0&1
	\end{array}
	\right]
	\left[
	\begin{array}{c}
		\mathbf{g}\\1
	\end{array}
	\right]\trans{\mathbf{X}}+\mathbb{Z}, \\
%	&=\left[\mathsf{H},\mathbf{d}\right]\cdot
%	\mathsf{diag}\left\{\exp\left(j\tilde{\bm{\Theta}}\right)\right\}\cdot
%	\tilde{\mathbf{g}}\trans{\mathbf{X}}+\mathbb{Z} \\
%	&=\left[\mathsf{H},\mathbf{d}\right]\cdot
%	\mathsf{diag}\left\{\tilde{\mathbf{g}}\circ\exp\left(j\tilde{\bm{\Theta}}\right)\right\}\cdot
%	\trans{\mathbf{X}}+\mathbb{Z} \\ 
	&=\left[\mathsf{H},\mathbf{d}\right]\cdot
	\mathsf{diag}\left\{\tilde{\mathbf{g}}\right\}\cdot
	\exp(j\tilde{\bm{\Theta}})\cdot	\trans{\mathbf{X}}+\mathbb{Z} \\
	&=\tilde{\mathsf{H}}\cdot
	\exp(j\tilde{\bm{\Theta}})\cdot	\trans{\mathbf{X}}+\mathbb{Z},  \label{eq:standard_model1}
\end{flalign}
where the augmented channel matrix $\tilde{\mathsf{H}}=\left[\mathsf{H},\mathbf{d}\right]\cdot
\mathsf{diag}\left\{\tilde{\mathbf{g}}\right\}\in \mathbb{C}^{\nr \times (n+1)}$, $\trans{\tilde{\mathbf{g}}}=\left[\trans{\mathbf{g}},1\right]$, and $\trans{\tilde{\bm{\Theta}}}=\left[\trans{{\bm{\Theta}}},0\right]$. Note that the direct path herein acts like an extra reflective element with zero phase shift.

\subsubsection{Equivalence Between Direct Path and RIS Element}
With an assumption of a control link between the RIS and the transmitter, the above model can be further simplified by carefully setting phase shifts at the RIS side. The simplified model facilitates subsequent mathematical processing and builds a unified model for RIS-aided SIMO links with/without a direct path.

Without any loss, we can rewrite the input $\mathbf{X}$ as $\exp(j\check{\Theta}_{n+1})\check{\mathbf{X}}$, where $\check{\Theta}_{n+1}\in[0,2\pi]$ can be any random phase and $\check{\mathbf{X}}$ is a phase-rotated input also satisfying the quadratic constraint \eqref{eq:input_constraints}. Accordingly, the channel model~\eqref{eq:standard_model1} is represented as follows:
\begin{flalign}
	\mathbb{Y}
%	 &=\tilde{\mathsf{H}}\cdot
%	\exp(j\tilde{\bm{\Theta}})\cdot	\trans{\mathbf{X}}+\mathbb{Z}   \nonumber \\
	&=\tilde{\mathsf{H}}\cdot
	\exp(j\tilde{\bm{\Theta}})\cdot	\exp(j\Theta_{n+1})\trans{\check{\mathbf{X}}}+\mathbb{Z} \\
	&=\tilde{\mathsf{H}}\cdot
	\exp\left(j\left(\tilde{\bm{\Theta}}+\Theta_{n+1}\right)\right)\cdot\trans{\check{\mathbf{X}}}+\mathbb{Z} \\
	&=\tilde{\mathsf{H}}\cdot
	\exp\left(j\check{\bm{\Theta}}\right)\cdot\trans{\check{\mathbf{X}}}+\mathbb{Z},
	\label{eq:standard_model2}
\end{flalign}
where the augmented phase shift vector $\check{\bm{\Theta}}=\trans{\left(\check{\Theta}_1,\cdots,\check{\Theta}_n,\check{\Theta}_{n+1} \right)} \in [0,2\pi]^{n+1}$ with the first $n$ entries  $\check{{\Theta}}_i=\Theta_i+\check{\Theta}_{n+1} $ ($\text{mod }2\pi$) for $i\in [n]$. 

Notice that the expanded and shifted phase vector $\check{\bm{\Theta}}$ is allowed to randomly take any value on $[0,2\pi]^{n+1}$ and the one only constraint on the phase-rotated input $\check{\mathbf{X}}$ is $\expec{\hermi{\check{\mathbf{X}}}\check{\mathbf{X}}}\le L\mathcal{E}$. Hence, without any loss, our considered RIS-aided channel with a direct path can be regarded as a new SIMO channel aided by $n+1$ RIS elements and without direct path, say \textit{equivalence between the direct path and an extra RIS element}.
%\footnote{Unfortunately, we shall remainder the reader that such the simplification may fail in the general MIMO case.}
%.

\subsubsection{Dimension Reduction}\label{subsubsec:dimension_reduction}
For less complexity involved in transceiver design and reduced feedback payload of the channel state information, one can use the SVD to reduce the dimension of an arbitrary input-constrained linear AWGN channel without any information loss (see, e.g. \cite{limoserwangwigger20_1}). Denote the rank of the augmented channel matrix $\tilde{\mathsf{H}}$ by $\tau\in \mathbb{N}_{+}$, and clearly $\tau \le \min\left\{n+1,\nr\right\}$. The singular value decompositon of $\tilde{\mathsf{H}}$ is denoted by 
$$
\tilde{\mathsf{H}}=\mathsf{U}\cdot \mathsf{diag}\left\{\rho_1,\cdots,\rho_{\tau},\bm{0}_{n+1-\tau}\right\}\cdot \hermi{\mathsf{V}},
$$
where $\mathsf{U}\in\mathbb{C}^{\nr\times \nr}$ and $\mathsf{V}\in \mathbb{C}^{(n+1)\times (n+1)}$ are corresponding unitary matrices, and $\rho_1$, $\rho_2\cdots$, and $\rho_{\tau}$ denotes $\tau$ nonzero singular values of $\tilde{\mathbb{H}}$.
With perfect channel state information at the receiver, we linearly transform the output as
\begin{flalign}
	 \hermi{\mathsf{U}} \mathbb{Y} =&
	\begin{bmatrix}
		&\mathsf{diag}\left\{\rho_1,\cdots,\rho_{\tau}\right\}&&\bm{0}_{\tau\times(n+1-\tau)}\\
		&\bm{0}_{(\nr-\tau)\times \tau}&&\bm{0}_{(\nr- \tau)\times (n+1-\tau)}
	\end{bmatrix}
	\begin{bmatrix}
		\hermi{\mathsf{V}}_{ 1}\\
		\hermi{\mathsf{V}_{ 2}}
	\end{bmatrix}\nonumber\\
	&\cdot\exp\left(j\check{\bm{\Theta}}\right)\trans{\check{\mathbf{X}}}+ \hermi{\mathsf{U}}\mathbb{Z},
\end{flalign}
where the semi-unitary matrix $\mathsf{V}_1$ consists of the first $\tau$ column vectors of $\mathsf{V}$.
Discard the last $\nr-\tau$ row vectors including only AWGN, and then the $\tau$-dimensional valid channel output is\footnote{For an RIS with massive reflective elements, the element grouping strategy is often used to avoid independently and simultaneously controlling all elements. The simplified model \eqref{eq:equiv} also applies to such control by blocks.}
\begin{equation}\label{eq:equiv}
	\check{\mathbb{Y}}=\check{\mathsf{H}}\exp\left(j\check{\bm{\Theta}}\right)\trans{\check{\mathbf{X}}}+\check{\mathbb{Z}},
\end{equation}
where the equivalent channel matrix $
\check{\mathsf{H}}\triangleq\mathsf{diag}\left\{\rho_1,\cdots,\rho_{\tau}\right\}\cdot\hermi{\mathsf{V}_{ 1}} \in \mathbb{C}^{\tau \times (n+1)}
$ and the equivalent AWGN $\check{\mathbf{Z}}= [\check{Z}_{ij}]\in \mathbb{C}^{\tau \times L}$ with i.i.d. components $  \check{Z}_{ij} \sim \mathcal{CN}\left( 0,1   \right)$ due to the rotation-invariant property of AWGN.

We shall call the model \eqref{eq:equiv} as the \textit{equivalent full-row-rank and purely-reflective model}\footnote{For simplicity, one may operate with this model to design $\check{\bm{\Theta}}$ and $\check{\mathbf{X}}$. In practical configuration, the transmitted signal should compensate the phase shift at the virtual reflective element (mathematically converted by a direct path) at the transmit antenna as $\mathbf{X}=\check{{\Theta}}_{n+1}\check{\mathbf{X}}$, while phase shifts at $n$ real reflective elements should configured as $\Theta_i=\check{\Theta}_i-\check{{\Theta}}_{n+1}$ (mod $2\pi$).}. Unless specially emphasized, the remainder of this paper is concerned with the equivalent full-row-rank and purely-reflective model.

%, and the capacity function $\mathsf{C}\left(\mathsf{H},\mathbf{g},\mathbf{d},\mathcal{E} \right) $ is therefore simply denoted by $\mathsf{C}\left(\check{\mathsf{H}},\mathcal{E} \right) $ as well. 

\subsection{Problem Formulation}

In this paper, we mainly focus on the fundamental limit to the RIS-aided SIMO channel under a quadratic cost constraint, over which the information is modulated on both the transmitted symbol from the single transmit antenna and the adjustable phase pattern at the RIS, i.e. joint transmission. With perfect channel state information at the transmitter and the receiver, the capacity of our considered channel is given as 
\begin{equation}\label{eq:capacity}
	\mathsf{C}\left(\check{\mathsf{H}},\mathcal{E} \right) 
	=\sup_{\mathbb{E}\left[\hermi{\check{\mathbf{X}}}\check{\mathbf{X}}\right]\le L \mathcal{E} \atop \supp \check{\bm{\Theta}} \subseteq [0,2\pi]^{n+1} } \frac{1}{L} \II\left(\check{\mathbf{X}},\check{\bm{\Theta}};\check{\mathbb{Y}}\right) ,
\end{equation}
where the constant $\mathcal{E}$ is the maximum allowed average transmit power.

We would like to point out that the joint transmission includes the most frequent usage of the RIS as a beamformer. In the beamforming mode, the phase shifts at the RIS side are usually prescribed and fixed until the channel state changes. Despite of an enhanced SNR level, the induced channel is still a SIMO channel, and hence equivalent to a scalar one with a DoF of one.  Optimization of those fixed phase shifts $\check{\bm{\Theta}}$ can be formulated into the following quadratic programming problem\footnote{Note that the involved problem \eqref{prob:UQP} is exactly a unit-modulus quadratic programming (UQP) problem since the Gram matrix $\check{\mathsf{H}}^\dag \check{\mathsf{H}}$ is positive semidefinite. Solving this UQP problem is out of the scope in this paper, and the reader who interested in relevant numerical methods please refer to \cite{tranter2017fast}. Although the UQP problem in a general form is non-convex and NP-hard due to the unit-modulus constraints, a popular relaxation technique known as semi-definite relaxation can be readily applied. Besides, it can be verified the optimal value remains unchanged if we place the constraint $\phi_{n+1}\in [0,2\pi]$ by $\phi_{n+1}=0$. }
\begin{flalign}\label{prob:UQP}
	\bm{\phi}^{\star}=	\argmax_{\check{\bm{\theta}}\in [0,2\pi]^{n+1}}
	 \left\|\check{\mathsf{H}}\cdot
	\exp\left(j\check{\bm{\theta}}\right) \right\|_2^2.
\end{flalign}
Denote the optimal value of the problem \eqref{prob:UQP} by ${F}^{\star}$. Then the maximized achievable rate in the beamforming mode is
\begin{flalign}\label{eq:AR_beamforming}
\bar{\const{R}}_{\text{bf}}(\check{\mathsf{H}}) = \log\left(1+ {F}^{\star} \mathcal{E} \right).
\end{flalign}

\section{Capacity Results for RIS-Aided SIMO Channels}\label{sec:capacity_results}
In this section, some capacity results, including the low-SNR asymptotics and the exact capacity in the rank-one case, for RIS-aided SIMO AWGN channels are presented.

\subsection{Upper Bounds}
In this subsection, two capacity upper bounds are presented by upper-bounding the trace of the covariance matrix of the equivalent input. 

We begin with a short introduction to the tool used in our derived capacity upper bounds. Consider an $m$-dimensional linear vector Gaussian channel
\begin{flalign}\label{eq:linear_channel}
	\mathbf{W}=\sqrt{\mathcal{E}}\mathsf{A}\mathbf{T}+ \mathbf{Z},
\end{flalign}
where $\mathsf{A}$ denotes the channel matrix, and the AWGN $\mathbf{Z}\sim \mathcal{CN}\left( \bm{0}_m,\mathsf{I}_{m}\right)$ is independent of the $\nt$-dimensional channel input $\mathbf{T}$. For a clear description, we define the equivalent input for the linear vector channel \eqref{eq:linear_channel} as $\bar{\mathbf{T}}\triangleq\mathsf{A}\mathbf{T}$. Since $\mathbf{T} \leftrightarrow \bar{\mathbf{T}} \leftrightarrow \mathbf{W}$ forms a Markov chain, then we have $\II\left(\mathbf{W}; \mathbf{T}\right)=\II\left(\mathbf{W};\bar{\mathbf{T}}\right)$.
The following lemma presents a capacity upper bound for the linear vector channel \eqref{eq:linear_channel} when the covariance matrix of $\bar{\mathbf{T}}$ has a fixed trace.
\begin{lemma}\label{lemma1}
	Let $\bar{\mathbf{T}} \in \mathbb{C}^{m}$ be a complex-valued random vector with the covariance matrix $\cov{\bar{\mathbf{T}}}$ 
	\begin{eqnarray}
		\cov{\bar{\mathbf{T}}} \triangleq \expec{\left( \bar{\mathbf{T}}-\expec{\bar{\mathbf{T}}}   \right)\hermi{\left( \bar{\mathbf{T}}-\expec{\bar{\mathbf{T}}}   \right)}}.
	\end{eqnarray}
Define $\kappa_{\scriptscriptstyle \bar{\mathbf{T}}} \triangleq \trace{\cov{\bar{\mathbf{T}}}}$. Then the mutual information between $\mathbf{W}$ and $\bar{\mathbf{T}}$ satisfies
	\begin{flalign}\label{eq:lemma2}
		\II\left(\mathbf{W};\bar{\mathbf{T}}\right) \le  m\log\left(1+\frac{\mathcal{E}}{m}\kappa_{\scriptscriptstyle \bar{\mathbf{T}}}\right),
	\end{flalign}
	where the upper bound is achieved by using circularly symmetric complex Gaussian random vector $\bar{\mathbf{T}}\sim \mathcal{CN}\left( \bm{0}_m, \frac{1}{m}\kappa_{\scriptscriptstyle \bar{\mathbf{T}}} \mathsf{I}_m \right)$.
\end{lemma}
\begin{IEEEproof}
	The above lemma for real-valued channels can be directly obtained by Eqs. (237)-(241) in \cite{limoserwangwigger20_1}. For the sake of concreteness, a proof of Lemma \ref{lemma1} for the complex-valued case is also provided in Appendix \ref{app:proof_lemma1}. 
\end{IEEEproof}

Then, based on Lemma~\ref{lemma1}, we investigate the capacity upper bound of our considered channel \eqref{eq:standard_model2}, whose output can be vectorized as follows
\begin{flalign}
	\mathbf{Y}=\mathsf{vec}\left(\check{\mathbb{Y}} \right) =\sqrt{\mathcal{E}}\bar{\mathbf{T}}+ \mathbf{Z},
\end{flalign}
and the corresponding equivalent input  
\begin{flalign}
	\bar{\mathbf{T}} &= \mathsf{diag}\left\{
	\check{\mathsf{H}}\cdot
	\exp\left(j\check{\bm{\Theta}}\right),\cdots,\check{\mathsf{H}}\cdot
	\exp\left(j\check{\bm{\Theta}}\right)
	\right\}\cdot\check{\mathbf{X}}/\sqrt{\mathcal{E}} \nonumber \\
	&= \frac{1}{\sqrt{\mathcal{E}}} \check{\mathbf{X}} \otimes \left(\mathsf{H}\cdot
\exp\left(j\check{\bm{\Theta}}\right) \right)
\end{flalign}
is an $(\nr L)$-dimensional complex-valued random vector. Then an upper bound on the trace of the covariance matrix of $\bar{\mathbf{T}}$ is given by the following Lemma.

\begin{lemma}\label{lemma2}
The trace $\trace{\cov{\bar{\mathbf{T}}}}$ is upper-bounded by
\begin{flalign}\label{eq:trace_bound}
	 \trace{\cov{\bar{\mathbf{T}}}} \le  L F^{\star} ,
\end{flalign}
where $F^{\star}$ is the optimal value of the quadratic programming problem \eqref{prob:UQP}.
\end{lemma}
\begin{IEEEproof}
	The lemma is concluded by noticing that
	\begin{flalign}
		& \trace{\cov{\bar{\mathbf{T}}}}  \nonumber \\
		=& \, \expec{\hermi{\left( \bar{\mathbf{T}}-\expec{\bar{\mathbf{T}}}   \right)} \left( \bar{\mathbf{T}}-\expec{\bar{\mathbf{T}}}   \right)} 
		\label{eq:trace2norm}
		\\
		=&\,	\expec{ \left\| \bar{\mathbf{T}} \right\|_2^2}  - \left\|   \expec{\bar{\mathbf{T}}} \right\|_2^2 \label{eq:covariance_power}\\
		\le & \,\expec{ \left\| \bar{\mathbf{T}} \right\|_2^2} \nonumber \\
		=&\,\expec{
			\sum_{\ell=1}^{L}	
			\left| \check{X}_{\ell} \right|^2 \cdot \left\| \check{\mathsf{H}}\exp(j\check{\bm{\Theta}}) \right\|_2^2/\mathcal{E}} \\
		\leq&\,   \expec{
			\sum_{\ell=1}^{L}	
			\left| \check{X}_{\ell} \right|^2/\mathcal{E}} \cdot \left( \max_{\bm{\phi} \in [0,2\pi]^{n+1}} \left\| \check{\mathsf{H}}\exp(j\bm{\phi}) \right\|_2^2 \right) \\
		\le&\, LF^{\star} \label{prob:norm_max},
	\end{flalign}
	where Eq. \eqref{eq:trace2norm} \rev{follows from} the linearity and cyclic invariance of the trace operator.
\end{IEEEproof}

Combining Lemma \ref{lemma1}, Lemma \ref{lemma2} and Eq. \eqref{eq:capacity} immediately leads to the following capacity upper bound.
\begin{theorem}[Maximum-Trace Based Upper Bound]\label{thm1:trace_ubd}
For the RIS-aided SIMO AWGN channel, the capacity $\mathsf{C}\left(\check{\mathsf{H}},\mathcal{E} \right)$ is upper-bounded by
	\begin{flalign}\label{eq:upper_bound}
		\mathsf{C}\left(\check{\mathsf{H}},\mathcal{E} \right)
		\le \tau \log\left(1+\frac{1}{\tau}\mathcal{E} F^{\star}\right). 
	\end{flalign}
\end{theorem}

Due to computation complexity involved in solving the UQP problem \eqref{prob:UQP}, we also provide a simpler capacity upper bound as follows.

\begin{proposition}[Frobenius-Norm Based Upper Bound]
For the RIS-aided SIMO AWGN channel, the capacity $\mathsf{C}\left(\check{\mathsf{H}},\mathcal{E} \right)$ is upper-bounded by
\begin{flalign}\label{eq:upper_bound2}
	\mathsf{C}\left(\check{\mathsf{H}},\mathcal{E} \right) 
	\le \tau \log\left(1+\frac{n+1}{\tau}\mathcal{E} \left\|\mathsf{H}\right\|_{\text{F}}^2\right). 
\end{flalign}
\end{proposition}
\quad\quad\emph{Sketch of Proof:}
The above bound can be easily derived by combining the definition of induced norm and the fact that any complex vector entry-wisely with a unit $\ell_2$ norm lies in a complex sphere. \QED

In the following two subsections, the low-SNR capacity slope for general cases and the exact capacity for rank-one channels are characterized \rev{by showing that there exists a signaling scheme has an achievable rate (asymptotically) coinciding with the upper bound} \eqref{eq:upper_bound}.

\subsection{Low-SNR Asymptotic}
For the linear vector Gaussian channel \eqref{eq:linear_channel}, the low-SNR asymptotic behavior of the mutual information is studied well. The first and the second-order asymptotics are given in the following lemma.

\begin{lemma}[\cite{prelovverdu04_1}]\label{prop:second_order_I}
	For any real number $\delta>\delta_0$
	\begin{flalign}\label{eqn:low_snr_asymptotic_conditions}
		\mathrm{Pr}\left\{ \left\| \mathbf{T} \right\|>\delta \right\} \le \exp\left(   -\delta^{\nu} \right),	
	\end{flalign}
	where $\delta_0>0$ and $\nu>0$ are positive constants. Then
	\begin{flalign}\label{eq:lowSNR}
		\II\left( \mathbf{T};\mathbf{W}  \right) = \log(e)\trace{\cov{\bar{\mathbf{T}}}} \mathcal{E}
		-	\frac{	\log(e)}{2}\trace{\cov{\bar{\mathbf{T}}}^2}\mathcal{E}^2+o\left( \mathcal{E}^2  \right)
	\end{flalign}	
	in units of bits per channel use (bpcu).
\end{lemma}

Before formally presenting the low-SNR capacity asymptotics, we define the following beamformed signaling that is shown to be \rev{asymptotically optimal at low SNR}.

\begin{definition}
	[Phase-Matched Bipolar Input] The joint channel input $\left( \mathbf{X}^{\star}, \bm{\Theta}^{\star}\right)$ is called the \textit{phase-matched bipolar input} for the channel \eqref{eq:standard_model2}, if $X_1^{\star},\cdots,X_{L}^{\star}$ are i.i.d. Bernoulli random variables taking on $\sqrt{\mathcal{E}}$ and $-\sqrt{\mathcal{E}}$ with equal probability and the RIS has a constant phase pattern as $\mathrm{Pr}\left\{\bm{\Theta}^{\star} = \bm{\phi}^{\star}\right\}=1$, i.e., acting as an optimal beamformer.
\end{definition}

Then the low-SNR capacity asymptotics is derived for the RIS-aided SIMO AWGN channel.
\begin{proposition} \label{prop:low_snr_asymptotic}
	The low-SNR capacity slope is given by
	\begin{flalign}
		\lim_{\mathcal{E}\to 0^{+}}	\mathsf{C}\left(\check{\mathsf{H}},\mathcal{E} \right)	/\mathcal{E}=\log(e)F^{\star}
	\end{flalign}
	bits per channel use per unit cost,
	which can be achieved by the phase-matched bipolar input.
\end{proposition}
\begin{IEEEproof}
	Let $B_1,\cdots ,B_{L}$ be i.i.d. Bernoulli random variables which take values on $\pm 1$ with equal probability. It can be verified that, by using the phase-matched bipolar input $\left( \mathbf{X}^{\star}, \bm{\Theta}^{\star}\right)$, the induced equivalent input (in a vector form) is $\bar{\mathbf{T}}^{\star}=\trans{\left(B_1,\cdots,B_{L}\right)}\otimes \left( \check{\mathsf{H}}\exp\left(j\bm{\phi}^{\star}\right) \right)$, whose covariance matrix has a trace of
	$ L F^{\star}$. Then, by using \eqref{eq:lowSNR}, we have
	\begin{flalign}
		&\frac{1}{L} \II\left(\mathbf{X}^{\star}, \bm{\Theta}^{\star};\mathbb{Y}\right)
		\nonumber \\
		=&\frac{1}{L} \II\left(\bar{\mathbf{T}}^{\star};\sqrt{\mathcal{E}}\bar{\mathbf{T}}^{\star}+\mathsf{vec}\left( \mathbb{Z} \right) \right) \nonumber \\
		=&\frac{\log(e)}{L}\left(
		L \mathcal{E} F^{\star} + o(\mathcal{E})
		\right) \nonumber \\
		=& \log(e)\mathcal{E} F^{\star} +o(\mathcal{E}),
	\end{flalign} 
which leads to $\liminf_{\mathcal{E}\to 0^{+}}	\mathsf{C}\left(\check{\mathsf{H}},\mathcal{E} \right)\ge \log(e)F^{\star}$. The converse part $\limsup_{\mathcal{E}\to 0^{+}}	\mathsf{C}\left(\check{\mathsf{H}},\mathcal{E} \right)\le \log(e)F^{\star}$ can be directly obtained from Theorem \ref{thm1:trace_ubd}.	 
\end{IEEEproof}

Using the phase-matched bipolar input, the induced binary-input vector channel is equivalent to a 2ASK-input real-valued scalar Gaussian channel
$$
Y=\sqrt{\mathcal{E}}\left\| \check{\mathsf{H}}\exp(j\bm{\phi}^{\star}) \right\|_2 B+N
=\sqrt{\mathcal{E}F^{\star}}B+N\rev{,}
$$
where the constellation point $B$ is uniformly distributed on $\left\{  -1 ,  +1\right\}$ and the AWGN $N\sim\mathcal{N}\left(0,\frac{1}{2}\right)$.

\begin{remark}
	[Generation to MIMO Phase-Modulated Channels] 
	By using same approach in Lemma~\ref{lemma2}, it can be easily verified that, for the MIMO phase-modulated channel $ \mathbf{Y} = \check{\mathsf{H}} \exp\left(j \check{\bm{\Theta}}\right) +\mathbf{Z}$, the optimal input at low SNRs is exactly a uniform distribution on the binary set $\left\{ \bm{\phi}^{\star},\bm{\phi}^{\star}+\pi\bm{1}_{n+1}\right\}$.
	\hfill\remarksymbol
\end{remark}

\subsection{Capacity of Rank-One Channel}\label{sec:rank-one}
In this subsection, we specially consider the rank-one case (i.e., $\tau=1$), which is motivated by the reasons: 1) in practical applications, the channel between the RIS and the receiver may be highly spatially correlated due to narrowly spaced antennas or reflective elements \cite{kilinc2021modeling}; 2) The case that the direct path between the transmit antenna and the receive antennas is much stronger than the reflective paths can be well-approximated by the rank-one channel. By using techniques introduced in Sec.~\ref{subsubsec:dimension_reduction}, the corresponding output of rank-one channel is given by 
\begin{flalign}\label{eq:rank_one_channel}
\trans{\mathbf{Y}}=\trans{\mathbf{h}}\cdot	\exp\left(j\check{\bm{\Theta}}\right)\cdot\trans{\check{\mathbf{X}}}+\trans{\mathbf{Z}},
\end{flalign}
where the channel state vector $\mathbf{h}=\trans{\left(h_1,\cdots,h_{n+1} \right)}\in \mathbb{C}^{n+1}$. 

%the number $\rho$ is the only nonzero singular value of the argumented channel matrix $\tilde{\mathbb{H}}$, and the $(n+1)$-dimensional complex-valued vector $\mathbf{v}_1$ is the eigenvector of $\hermi{\check{\mathsf{H}}}\check{\mathsf{H}}$ corresponding to the only nonzero eigenvalue $\rho_1^2$.

In the rank-one case, it can be easily seen that the optimal value of the optimization problem \eqref{prob:UQP} has a closed form as follows
\begin{flalign}
\sqrt{ F^{\star}}
%&=
%  \max_{\bm{\phi} \in [0,2\pi]^{n+1}}
% \left| \trans{\mathbf{h}} \exp(j\bm{\phi}) \right| \nonumber \\
 &= \max_{\bm{\phi} \in [0,2\pi]^{n+1}} \left|\sum_{i=1}^{n+1}h_i\exp(j\phi_i)  \right|\nonumber \\
 &=  \left\| \mathbf{h} \right\|_1, \label{eq:rank_one_norm}
\end{flalign}
and the optimal solution to \eqref{prob:UQP} is simply given by $\phi_i =-\angle{h_i}$, for $i\in [n+1]$, where \eqref{eq:rank_one_norm} follows from the absolute value inequality. Hence, by fixing all phase shifts $\Phi_i =-\angle{h_i}$, the above rank-one channel \eqref{eq:rank_one_channel} can be reduced to a quadratically constrained vector Gaussian one
\begin{flalign}
	{\mathbf{Y}}=\left\| \mathbf{h} \right\|_1\check{\mathbf{X}}+\mathbf{Z}.
\end{flalign}
For the above channel, i.i.d. circularly symmetric complex Gaussian codebook achieves the maximum rate $ \log\left(1+\left\| \mathbf{h} \right\|_1 \mathcal{E} \right)$ (averaged over $L$ channel uses), which coincides with the capacity upper bound \eqref{eq:upper_bound} in Theorem \ref{thm1:trace_ubd}. For this reason, we conclude that our considered channel in the rank-one case has a closed-form capacity formula as
\begin{flalign}
	\mathsf{C}\left(\check{\mathsf{H}},\mathcal{E} \right) = \log\left(1+\left\| \mathbf{h} \right\|_1^2 \mathcal{E} \right), ~\text{if } r=1.
\end{flalign}

To summarize, \textit{in low-SNR or rank-one channels, the optimal configuration of the RIS is beamforming, i.e., modulating extra information on phase shifts at the RIS side brings no achievable rate gain}. In such scenarios, increasing the quantization bits of RIS phase shifts, rather than re-configuration rate, may be helpful. Additionally, standard modulation and coding techniques for scalar Gaussian channels can be readily used to approach the capacity in those cases.
%%
%%improvement on the system performance, which heavily relaxes the requirement of a real-time control link between the RIS and the transmitter. 
%\hfill\remarksymbol \end{remark}

\section{Hypersphere-Modulated Vector Gaussian Channel}\label{sec:hp}
In this section, a class of channels with input constrained on a multidimensional sphere is investigated. Those results will play a vital role in analyzing the novel transceiver architecture proposed in Sec.~\ref{sec:qr-sic_transceiver}.

For both generality of the results and mathematical convenience, \rev{we are concerned with the real-valued case, while the commonly considered complex-valued case is equivalent to the real-valued case with even dimensions.} Let $m$ be an integer no less than $2$, and then we define the \textit{hypersphere-modulated vector Gaussian channel} (HSM-VGC) as the following multidimensional AWGN channel with a hyperspherical support constraint 
\begin{flalign}
	\mathbf{Y}=\sqrt{\mathsf{snr}}\cdot \mathbf{X}+\mathbf{N},
\end{flalign}
where
% the $m$-dimensional real-valued random vector $\mathbf{Y}$ denotes the channel output,
  the $m$-dimensional real-valued random vector $\mathbf{X}$ is the channel input subject to a hyperspherical support constraint 
\begin{flalign}
	\mathbf{X}\in \sqrt{m}\mathcal{S}_{m-1},
\end{flalign}
$\mathcal{S}_{m-1}$ denotes the $(m-1)$-dimensional unit hypersphere in the $m$-space, i.e.,
\begin{flalign}
\mathcal{S}_{m-1}\triangleq \left\{ \mathbf{x}\in \mathbb{R}^{m} \vert \trans{\mathbf{x}}\mathbf{x} =1 \right\},
\end{flalign}
and the real-valued AWGN $\mathbf{N}\sim \mathcal{N}(\bm{0}_{m}, \mathsf{I}_{m})$.
We denote the capacity of the HSM-VGC by 
\begin{flalign}
	\const{C}_{\textrm{S}}^{(m)}(\mathsf{snr})
	\triangleq  \sup_{\mathsf{supp}\,\mathbf{X}\subseteq \sqrt{m}\mathcal{S}_{m-1}} \II\left(\mathbf{X};\sqrt{\mathsf{snr}}\cdot \mathbf{X}+\mathbf{N}\right) .
\end{flalign}	

The complex-valued HSM-VGC (equivalently regarded as the case of even $m$) has been independently investigated in \cite{sedaghat2016cpm,karout2017ofc}, where the uniform distribution on the support is proved to be capacity-achieving at any SNR and an integral-based capacity expression is presented. Additionally, in the special case of $m=2$, the HSM-VGC is degenerated to the well-know phase-modulated channel, where the information is modulated on the phase of a complex number. In \cite{wyner66_1}, the low- and high-SNR capacity asymptotics and bounds on the cardinality of a polyphase codebook with a prescribed minimum distance are given for phase-modulated channels. Besides, it is well-known that a random spherical codebook of infinite blocklength is capacity-achieving for the scalar real-valued AWGN channel. Hence, the capacity of the HSM-VGC per dimension (i.e., $\const{C}_{\textrm{S}}^{(m)}(\mathsf{snr})/m$) converges to $\frac{1}{2}\log(1+\snr)$ as the dimension $m\to +\infty$. Nevertheless, a comprehensive analysis of channel capacities for HSM-VGCs of an arbitrarily finite dimension is less investigated.

%	well good approximation, accurate characterization of high-SNR behaviors.} Another issue 

In the following theorem, we generalizes existing results in \cite{wyner66_1,sedaghat2016cpm} to arbitrarily-dimensional Euclidean spaces and present a modified capacity expression as well as a simple alternative proof.

\begin{theorem}[\cite{wyner66_1,sedaghat2016cpm}] \label{thm:2}
	Let $T\sim \chi^2_{m}(m,\snr)$ be a noncentral chi-squared distributed random variable with $m$ degrees of freedom and the probability density function
		\begin{flalign}
			p_{T}(t)=&\frac{\mathsf{snr}}{2}\left(\frac{t}{m}\right)^{\frac{m}{4}-\frac{1}{2}}\exp\left(-\frac{\mathsf{snr}\cdot(t+m)}{2}\right)\nonumber \\
			&\cdot
			I_{\frac{m}{2}-1}(\mathsf{snr}\cdot\sqrt{m t}),~t\ge 0, \label{eq:pdf_chisquared}
		\end{flalign}
	where $I_v(\cdot)$ denotes the $v$-th order modified Bessel function of the first kind. Then the capacity (in bits per channel use) of the HSM-VGC is
	\begin{flalign}\label{eq:HSMVGC_capacity}
		&\const{C}_{\textrm{S}}^{(m)}(\mathsf{snr}) \nonumber \\
		=&\left(\frac{m}{2}-1\right)\log\frac{\snr}{2 }  +\frac{m-2}{4}\expec{\log mT}+\left(m\snr \right)\log(e) \nonumber \\
		&-\expec{\log I_{\frac{m}{2}-1}(\sqrt{mT}\snr)} -\log(\Gamma(m/2)) ,
	\end{flalign}
and can be achieved by the uniform distribution $\textrm{Unif}\left( \sqrt{m}\mathcal{S}_{m-1} \right)$, where $\Gamma(\cdot)$ denotes the Gamma function.
\end{theorem}
\begin{IEEEproof}
	See Appendix \ref{app:pre} for prerequisite properties of spherically symmetric distributions and Appendix \ref{app:proof_thm2} for the proof.
\end{IEEEproof}

Next we derive a higher-order capacity asymptotic for general HSM-VGCs, by which we can evaluate their capacities with remarkable accuracy.

\begin{theorem}\label{thm:3}
	The high-SNR capacity of the HSM-VGC is given as
\begin{flalign}\label{eq:asymptotics}
	\const{C}_{\textrm{S}}^{(m)}(\mathsf{snr})
	=&\left(\frac{m-1}{2}\right)\log\left( \frac{ m\snr}{2e}\right)+ \log\left( \frac{2\sqrt{\pi}}{\Gamma(m/2)} \right)\nonumber \\
	&+\left( \frac{m}{2}-\frac{7}{4}+\frac{5}{4m} \right)\frac{\log(e)}{\snr} +o\left(\frac{1}{\snr}\right).
\end{flalign}
\end{theorem}
\begin{IEEEproof}
 See Appendix \ref{app:proof_thm3}.
\end{IEEEproof}

For the completeness of the results, we also present the above-mentioned trivial upper bound on $\const{C}_{\textrm{S}}^{(m)}(\mathsf{snr})$ as well as its low-SNR slope without proof.
\begin{proposition}
	The capacity of the HSM-VGC is upper-bounded by 
	\begin{flalign}\label{eq:ubd_hsm-vgc}
		\const{C}_{\textrm{S}}^{(m)}(\mathsf{snr})
		\le \frac{m}{2}\log\left( 1+\snr \right),
	\end{flalign}
	and the low-SNR capacity slope is therefore given by
		\begin{flalign}
		\lim_{\snr\to 0^{+}}	\const{C}_{\textrm{S}}^{(m)}(\mathsf{snr})	/\snr=\left(m\log(e)\snr\right)/2.
	\end{flalign}
\end{proposition}

\begin{remark}
	[Coincidence With Existing Result]
	In \cite{wyner66_1}, A. D. Wyner has proved that
	\begin{flalign}\label{eq:wyner_asymptotic}
	\const{C}_{\textrm{S}}^{(2)}(\mathsf{snr})=\frac{1}{2}\log\left(\frac{4\pi}{e}\snr\right)+o(1),
\end{flalign}
	for a phase-modulated channel (i.e., the HSM-VGC with $m=2$). By using Theorem \ref{thm:3}, the decay rate of the remainder term can be further determined as
	\begin{flalign}\label{eq:our_asymptotic}
	\const{C}_{\textrm{S}}^{(2)}\left(\mathsf{snr}\right)=
	\frac{1}{2}\log\left(\frac{4\pi}{e} \snr\right) -\frac{\log(e)}{8\snr}
	+o\left(\frac{1}{\snr}\right).
\end{flalign}	
%\hfill\remarksymbol 
\end{remark}

For our considered RIS-aided channel, the above asymptotic results on phase-modulated channels and arbitrary-dimensional HSM-VGCs will be repeatedly used in the subsequent analysis of achievable rates.

\section{QR-SIC Transceiver With Complex Gaussian Codebook}\label{sec:qr-sic_transceiver}
In this section, we propose a transceiver architecture based on QR decomposition and successive interference cancellation (SIC) to extract extra information modulated on the phase shifts at the RIS side, for which achievable rates as well as their asymptotics are given by using capacity results on HSM-VGCs.

Denote the QR decomposition of the channel matrix $\check{\mathsf{H}}$ as 
$$
\check{\mathsf{H}}=\mathsf{Q}\mathsf{R},
$$
where $\mathsf{Q}$ is a $\tau \times \tau$ unitary matrix and $\mathsf{R}=[r_{ij}]\in \mathbb{C}^{\tau\times \left(n+1\right)}$ is an upper triangular matrix with nonnegative diagonal entries. Then we linearly transform the channel output as
\begin{flalign}
	\tilde{\mathbb{Y}}=\hermi{\mathsf{Q}}\mathbb{Y}=\mathsf{R}\cdot\exp\left(j\check{\bm{\Theta}}\right)\cdot\trans{\check{\mathbf{X}}}+\check{\mathbb{Z}},
\end{flalign}
which can be further regarded as $\tau$ vector sub-channels as follows 
\begin{flalign}
	\trans{\tilde{\mathbf{Y}}}_{i}=\sum_{k=i}^{n+1} r_{ik}\exp(j\check{\Theta}_k)\trans{\check{\mathbf{X}}}+\mathbf{Z}_i, ~\text{for } i\in [\tau].
\end{flalign} 
By decoding each sub-channel in a descending order and using successive interference cancellation (SIC), we propose a new transceiver architecture in what follows for joint transmission accomplished by the single transmit antenna and the RIS.

\subsection{Transceiver Architecture}

The following strategy called \textit{partially beamforming and partially information-carrying} (PB/PIC) is used in our QR-SIC approach.

\begin{itemize}
	\item \textit{1) Partially Beamforming:}
	We particularly let $n-\tau+2$ phase shifts $\check{\Theta}_{\tau},\check{\Theta}_{\tau+1},\cdots,\check{\Theta}_{n+1}$ be constant to facilitate us decode $\check{\mathbf{X}}$ from the $\tau$-th sub-channel\footnote{Another way to decode $\check{\mathbf{X}}$ is treating the channel \eqref{eq:equiv} as a SIMO phase-noisy channel, for which both the rate analysis and the modem design are complicated as compared to that for the standard AWGN channel.}. In this setting, the $\tau$-th sub-channel
	output is simply degraded to a quadratically constrained AWGN channel
	\begin{flalign}\label{eq:siso}
		\trans{\tilde{\mathbf{Y}}}_{\tau}=\left(\sum_{k=\tau}^{n+1} r_{\tau k}\exp(j\check{\theta}_k)\right)\trans{\check{\mathbf{X}}}+\trans{\mathbf{Z}}_{\tau}.
	\end{flalign}
		For notation simplicity, we define the constant $${\const{G}_{\tau}}\triangleq\sum_{k=\tau}^{n+1} \left|r_{\tau k}\right|. $$
	In line with the aforementioned result for the rank-one case (see Sec. \ref{sec:rank-one}), the channel coefficient $\left(\sum_{k=\tau}^{n+1} r_{\tau k}\exp(j\check{\theta}_k)\right)$ in \eqref{eq:siso} is maximized as $\const{G}_{\tau}$ by letting $\check{\theta}_{k}=-\angle{r_{\tau k}}$ for $k=\tau, \cdots, n+1$, and hence, the induced vector channel \eqref{eq:siso} is
	\begin{flalign}\label{eq:siso3}
		\trans{\tilde{\mathbf{Y}}}_{\tau}=
		\const{G}_{\tau}
		\trans{\check{\mathbf{X}}}+\trans{\mathbf{Z}}_{\tau}.
	\end{flalign}
	\item \textit{2) Partially Information-Carrying: }
	For a higher achievable rate, $\tau-1$ streams of extra information are individually modulated on the remainder phase shifts $\check{\Theta}_1,\cdots,\check{\Theta}_{\tau-1}$. Due to the optimality of uniformly-distributed phase for phase-modulated channels at any SNR, we let $\check{\Theta}_1,\cdots,\check{\Theta}_{\tau-1}$ be independently and uniformly distributed on $[0,2\pi]$, say \textit{continuous phase shift keying} (CPSK) signaling. Then random phase shifts $\check{\Theta}_{\tau-1},\cdots,\check{\Theta}_1$ are successively decoded  by using SIC from the $(\tau-1)$-th sub-channel to the first one. In more details, with the knowledge of beamforming phase shifts $\check{\theta}_{\tau},\cdots, \check{\theta}_{n+1}$, known $\check{\mathbf{X}}$ decoded by the $\tau$-th sub-channel, and known (but random) phase shifts $\check{\Theta}_{\tau-1},\cdots,\check{\Theta}_{i+1}$ individually extracted from the $(\tau-1)$-th sub-channel to the $(i+1)$-th one, the $i$-th sub-channel, $i\in [\tau-1]$, can be converted into (with a little abuse of notation) the following phase-modulated channel
	\begin{flalign}
		 &~\trans{\tilde{\mathbf{Y}}}_{\!i}=\left(\sum_{k=i}^{\tau-1} r_{ik} \exp(j\Theta_k)+\sum_{k=\tau}^{n+1} \left|r_{ik}\right|
		\right)\trans{\check{\mathbf{X}}} +\trans{\mathbf{Z}}_{i} \nonumber \\
		\Longleftrightarrow &~ \trans{\tilde{\mathbf{Y}}}_{\!i}=r_{ii}\exp(j\Theta_i)\trans{\check{\mathbf{x}}}+\trans{\mathbf{Z}}_{i} \\
		\Longleftrightarrow &~ \tilde{{Y}}_{i}=r_{ii}\left\|\check{\mathbf{x}}\right\|_2\exp(j\Theta_i)+Z_{i}, \label{eq:phase-only}
	\end{flalign}
where \eqref{eq:phase-only} is obtained by treating the known realization $\left\|\check{\mathbf{x}}\right\|_2$ of the channel input $\check{\mathbf{X}}$ as a channel state vector and using the techniques in Sec.~\ref{subsubsec:dimension_reduction} to transform a SIMO channel to a SISO one.
\end{itemize}

Clearly, the partially beamforming strategy is used to enhance the SNR of the last sub-channel for a better recovery of the transmitted signal $\mathbf{X}$, while the partially information-carrying strategy is used for an increased system throughput. 

\rev{In the following two subsection, achievable rate analysis of our proposed transceiver is carried out under different assumption of joint inputs.}

\subsection{Achievable Rates by Gaussian/CPSK Input}

It is well-known that, for the sub-channel model \eqref{eq:siso3}, using the i.i.d. input does not induce any rate loss, and hence, the channel model \eqref{eq:siso3} in a multi-letter form is converted into the following scalar complex-valued AWGN channel
\begin{flalign}\label{eq:siso2}
	\tilde{{Y}}_{\tau}=\const{G}_{\tau}\check{X}+{Z}_i,
\end{flalign}
for which the complex Gaussian input $\check{X}\sim \mathcal{CN}( 0 ,\mathcal{E})$ achieves the capacity $	\log \left( 1+\const{G}_{\tau}^2 \mathcal{E} \right)$.

For \rev{the above} reason, in this subsection, we investigate the achievable rates of our proposed QR-SIC transceiver when using i.i.d. complex circularly symmetric Gaussian input $\check{\mathbf{X}}\sim \mathcal{CN}(\bm{0}_{L},\mathcal{E}\mathsf{I}_{L})$. In the proposed transceiver architecture, the extra information is modulated on $\tau-1$ uniformly distributed phase shifts and then decoded by all sub-channels except for the last one. The achievable rate that successfully extracting $\check{\Theta}_i$ from the $i$-th sub-channel is given by
	\begin{flalign}
		&\sup_{\Theta_i}  \II\left(\tilde{\mathbf{Y}}_{i};\Theta_i \vert \check{\mathbf{X}},\Theta_{i+1},\cdots,\Theta_{n+1} \right) \nonumber \\
		= 	& \, \sup_{\Theta_i} \II\left(r_{ii}\left\|\check{\mathbf{x}}\right\|_2\exp(j\Theta_i)+Z_{i}; \Theta_i \right) \label{eq:vector2scalar} \\
		=& \,	\const{C}_{\textrm{S}}^{(2)}\left(r_{ii}^2 \left\|\check{\mathbf{x}}\right\|_2^2 \right) ,\label{eq:phase2circle}
	\end{flalign}
for any $i\in [\tau-1]$, where Eq.~\eqref{eq:vector2scalar} is obtained by Eq.~\eqref{eq:phase-only}, and Eq.~\eqref{eq:phase2circle} follows from the equivalence between a phase-modulated channel and a $2$-dimensional HSM-VGC. Note that the instantaneous SNR (or say, channel state) $r_{ii}^2 \left\|\check{\mathbf{x}}\right\|_2^2$ varies with the instantaneous input $\check{\mathbf{x}}$ from the transmit antenna. For such \rev{a} channel, the maximum achievable rate is an averaged capacity over all channel states since the information of channel states is available at the transmitters \cite{shannon58_1}.

Hence, the achievable rate of the QR-SIC transceiver with complex Gaussian and CPSK input can be calculated as shown in the following proposition.

\begin{proposition}
	[Achievable Rates]
	Let $\hat{T}\sim \chi^2_{2L}$ be a central chi-squared distributed random variable with $2L$ degrees of freedom and the probability density function
\begin{flalign}
		p_{  \hat{T}}(t)=&\frac{ 1}{2^L\cdot (L-1)!}t^{L-1}\exp\left(-\frac{t}{2}\right),~t\ge 0. \label{eq:pdf_chisquared2}
	\end{flalign}
Then the achievable rate of the QR-SCI transceiver using Gaussian/CPSK input is
	\begin{flalign}\label{eq:rate_1}
		\const{R}_{\text{QR-SIC,\,\rmnum{1}}} =
		\log \left( 1+ \const{G}_{\tau}^2 \mathcal{E} \right) + \frac{1}{L}\sum_{i=1}^{\tau-1} \expec{ \const{C}_{\textrm{S}}^{(2)}\left(\frac{r_{ii}^2 \mathcal{E} \hat{T}}{2} \right) }.
	\end{flalign}
\end{proposition}
\begin{IEEEproof}
	Note that
	\begin{flalign}
	\const{R}_{\text{QR-SIC,\,\rmnum{1}}}=&
	\log \left( 1+ \const{G}_{\tau}^2 \mathcal{E} \right) \nonumber \\ &+\frac{1}{L}\sum_{i=1}^{\tau-1}\mathbb{E}_{\check{\mathbf{X}}\sim \mathcal{CN}(\bm{0}_{L},\mathcal{E}\mathsf{I}_{L})}  \left[  \const{C}_{\textrm{S}}^{(2)}\left(r_{ii}^2  \left\|\check{\mathbf{X}}\right\|_2^2  \right) \right]  \nonumber
	\end{flalign}
		where the factor $\frac{1}{L}$ follows from the fact that a phase-modulated symbol occupies $L$ channel uses. Then we conclude the proposition by noticing that $\left\|\check{\mathbf{X}}\right\|_2^2$ equals (in distribution) the sum of $2L$ \rev{squares} of i.i.d. zero-mean real-valued Gaussian random variables with variance $\mathcal{E}/2$.
\end{IEEEproof}

Note that the accurate evaluation of the above rate involves complicated integral or Monte-Carlo simulation. To avoid this issue, we also characterize the high-SNR rate asymptotics in the following proposition, which is shown to be a good approximation later.

\begin{proposition}[Asymptotics]\label{prop5}
	Asymptotic achievable rate of the QR-SCI transceiver using Gaussian/CPSK input is
		\begin{flalign}
		\const{R}_{\text{QR-SIC,\,\rmnum{1}}}=&
		\log \left( 1+ \const{G}_{\tau}^2 \mathcal{E} \right) +\frac{1}{L}\sum_{i=1}^{\tau-1} \bar{\const{R}}_{\text{CPSK},i}
	\end{flalign}
	where the average rate $\bar{\const{R}}_{\text{CPSK},i}$ for $i\in[\tau-1]$ is given by
	\begin{flalign}
		&\bar{\const{R}}_{\text{CPSK},i} \nonumber \\
		=&
		\begin{cases}
			\frac{1}{2}\log\left(\frac{4\pi}{e} r_{ii}^2 \mathcal{E} \right) -\frac{1}{2} \log(e) \gamma +o(1),\,L=1;\\
			\frac{1}{2}\log\left(\frac{4\pi}{e} r_{ii}^2 \mathcal{E} \right) +\frac{1}{2} \log(e) \psi(L) -\frac{\log(e)}{8(L-1)r_{ii}^2 \mathcal{E} }+o\left( \frac{1}{\mathcal{E}} \right),\text{else,}
		\end{cases}
	\end{flalign}
	where $\gamma$ and $\psi(\cdot)$ denote the Euler constant and the diagamma function (see Eq. \eqref{eq:digamma}), respectively.
\end{proposition}
\begin{IEEEproof}
	See Appendix \ref{app:prop5}.
\end{IEEEproof}

\subsection{Achievable Rates by Hypersphere/CPSK Input}

%\begin{remark}
%	[Universal Coding]
%	\crh{Capacity-achieving distributions of phase-modulated channels at any SNR is uniformly distributed on $[0,2\pi]$.}
%\hfill\remarksymbol \end{remark}

As mentioned before, the channel states of $\tau-1$ phase-modulated subchannels \eqref{eq:phase-only} are determined by the channel input of \eqref{eq:siso2} and hence time-varying. A universal coding scheme may be needed to achieve ergodic capacities of those phase-modulated channels with channel fading, which may heavily increase the encoding complexity \cite{shannon58_1}. To remove this disadvantage, a natural way is letting the $\ell_2$-norm of instantaneous $\check{\mathbf{X}}$ be the constant $\sqrt{L\mathcal{E}}$, i.e., $\left\| \check{\mathbf{X}} \right\|_2 =\sqrt{L\mathcal{E}}$, so that phase-modulated channel \eqref{eq:phase-only}	\begin{flalign}
 \tilde{{Y}}_{i}=\sqrt{L\mathcal{E}}r_{ii}\exp(j\Theta_i)+Z_{i}
\end{flalign}
has a constant channel state.

Under the above norm constraint, the $L$-dimensional complex-valued vector channel \eqref{eq:siso3} is equivalent to the aforementioned $2L$-dimensional HSM-VGC with $\mathsf{snr}=\const{G}_{\tau}^{2}\mathcal{E}$, for which the joint distribution $\trans{\left(\rp{\check{\mathbf{X}}},\ip{\check{\mathbf{X}}} \right)}$ uniformly distributed on  $\sqrt{L\mathcal{E}}\mathcal{S}_{2L-1}$ achieves the capacity. For this reason, such a constant-norm input is called \textit{hypersphere signaling}. By using capacity expression of HSM-VGCs, the achievable rate of the QR-SIC transceiver with hypersphere/CPSK input can be immediately obtained as follows.

\begin{proposition}
	[Achievable Rates]
	The achievable rate of the QR-SCI transceiver using hypersphere/CPSK input is
	\begin{flalign}\label{eq:rate2}
	\mathsf{R}_{\text{QR-SIC,\,\rmnum{2}}}  = \frac{1}{L} \left(\const{C}_{\textrm{S}}^{(2L)}\left(\const{G}_{\tau}^2 \mathcal{E}  \right)+
		\sum_{i=1}^{\tau-1}\const{C}_{\textrm{S}}^{(2)}\left(r_{ii}^2 L\mathcal{E}  \right) \right).
	\end{flalign}
\end{proposition}

%Relying on Theorem \ref{thm:2}, numerical methods such as Monte Carlo method can be used to determine the achievable rate as accuracy as possible. Nevertheless, a high-order asymptotic is needed. 

Substituting Eq.~\eqref{eq:asymptotics} in Theorem \ref{thm:3} into Eq.~\eqref{eq:rate2}, we obtain the high-SNR asymptotics of $\mathsf{R}_{\text{QR-SIC,\,\rmnum{2}}}$ as follows.
	
\begin{corollary}\label{cor:2}
Asymptotic achievable rate of the QR-SCI transceiver using hypersphere/CPSK input is
	\begin{flalign}
		& \mathsf{R}_{\text{QR-SIC,\,\rmnum{2}}}  \nonumber \\
		=& \frac{2L-1}{2L} 
		\log\left(L  \const{G}_{\tau}^2 \mathcal{E}/e\right)+\frac{1}{2L}\sum_{i=1}^{\tau-1}  \log\left(\frac{4\pi}{e} r_{ii}^2 L\mathcal{E} \right)\nonumber \\
		&+ \frac{1}{L}\log\left( \frac{2\sqrt{\pi}}{(L-1)!} \right)+\left( 1-\frac{7}{4L}+\frac{5}{8L^2} \right)\frac{\log(e)}{\const{G}_{\tau}^2 \mathcal{E}} \nonumber \\
		&-\sum_{i=1}^{\tau-1} \left(  \frac{\log(e)}{8r_{ii}^2 L^2\mathcal{E} }  \right) + o\left( \frac{1}{\mathcal{E}}\right).
	\end{flalign}
\end{corollary}

\subsection{\rev{Summary}}
In this subsection, some important issues of our proposed transceiver are discussed.

 \textit{1) Achievable DoFs:} For a given signaling method, its achievable DoF is usually defined as the limiting pre-log factor of the achievable rate. Hence, for our proposed QR-SIC transceiver with two different inputs, the DoFs are given by
		\begin{flalign}
			\mathsf{DoF}_{\text{QR-SIC,\,\rmnum{1}/\rmnum{2}}} 
			\triangleq 
			\lim_{\mathcal{E}\to +\infty}  \frac{\const{R}_{\text{QR-SIC, \rmnum{1}/\rmnum{2}}}}{\log \mathcal{E}}.
		\end{flalign}
		From Proposition \ref{prop5} and Corollary \ref{cor:2}, it can be easily verified that the DoF of the QR-SIC transceiver with Gaussian/CPSK input is
		\begin{flalign}
 \mathsf{DoF}_{\text{QR-SIC,\,\rmnum{1}}} 
			= 1+\frac{\tau-1}{2L} .
		\end{flalign}
		while that with hypersphere/CPSK input is
		\begin{flalign}
			\mathsf{DoF}_{\text{QR-SIC,\,\rmnum{2}}} = 1+\frac{\tau-2}{2L} .
		\end{flalign}
The deficiency of hypersphere/CPSK input is caused by imposing an extra hypersphere support constraint on the channel \eqref{eq:siso3}.
Clearly, as compared to using the conventional SIMO channel with/without a beamforming RIS, extra DoFs of $\frac{\tau-1}{2L}$ or $\frac{\tau-2}{2L}$ can be extracted by our transceiver.  In \cite{cheng2021dof}, the authors have shown that, for \textit{almost all} rank-$\tau$ channels matrices $\check{\mathsf{H}}$ and the SRR $L=1$, the maximum DoF is $\min(1+ \frac{n}{2}, \tau)$ when a single RF transmitter and the $(n+1)$-element RIS (or equivalently $n$ reflective elements and a direct path) are used for joint transmission\footnote{Nevertheless, for some channels \textit{in a specific form}, the maximum DoF may varies from $1 + \frac{\tau-1}{2}$ to $\min(1+ \frac{n}{2}, \tau)$.}. Hence, the proposed QR-SIC transceiver using Gaussian/CPSK input achieves full DoFs when $n+1$ channel column vectors between each RIS element and all receive antennas are linearly independent, i.e., $\tau=n+1$. The main reason for potential DoF loss in general cases is the sub-optimal treatment on the involved MIMO phase-modulated channel. To overcome this issue, spatially correlation may be exploited for an increased achievable rate.

% \rev{For a small number of reflective elements such as $n+1\ge \tau$. Our scheme can achieve the maximum DoFs} 

\textit{2) Modulation and Coding Schemes:} It can be seen that our proposed QR-SIC transceiver is free of complicated treatment on the SIMO phase-noisy channel and the MIMO phase-modulated one, for which there is no simple capacity-achieving signaling scheme in general. Instead, classic coding and modulation schemes tailored for the scalar AWGN channel and phase-modulated one can be readily applied to the QR-SIC transceiver, so heavily simplifies the implementation complexity.

%. for which many existing modulation and coding schemes can be readily used to approach its channel capacity. 

% $\min\left\{ \frac{n}{2},\tau\right\}$. \begin{flalign}
%			& \II \left( \mathbf{Y}; X, \tilde{\bm{\Theta}} \right) \nonumber \\
%			=  & \II \left( \mathsf{H}\exp(j\tilde{\bm{\Theta}})X+\bm{Z}; X, \tilde{\bm{\Theta}} \right) \nonumber \\
%			= & \underbrace{\II \left( \mathsf{H}\exp(j\tilde{\bm{\Theta}})X+\bm{Z}; X \right)}_{\text{phase noisy channel}} + \underbrace{\II \left( \mathsf{H}\exp(j\tilde{\bm{\Theta}})X+\bm{Z}; \tilde{\bm{\Theta}} \vert X \right)}_{\text{phase-modulated channel}}\label{eq:channel_decomposition}
%		\end{flalign}
%		where the maximum DoF extracted from the phase noisy channel is no larger than $\frac{1}{2}$ due to the phase ambiguity\footnote{For phase-noisy channel, since the channel can be deemed as a fading channel. The maximum DoF or fading number is no larger than $1$. }, and for the phase-modulated channel
%		\begin{flalign}
%			\II \left( \mathsf{H}\exp(j\tilde{\bm{\Theta}})X+\bm{Z}; \tilde{\bm{\Theta}} \vert X \right) \le 
%			\II \left(\sqrt{\mathcal{E}} \mathsf{H}\exp(j\tilde{\bm{\Theta}})+\bm{Z}; \tilde{\bm{\Theta}}  \right)
%		\end{flalign}
%		due to the concavity of mutual information with respect to SNR.

%\textit{3) Required CSI:} RIS is a
%nearly-passive device, and hence it cannot process and transmit
%pilot signals. To account for this practical constraint. A control link between the RIS and the transmitter is needed.
%Low SNR case: RIS configuration need CSI.
%High-SNR case: QR-based scheme does not need.

\textit{3) Compatibility With Finite-Bit Phase Quantization:} Another advantage of the QR-SIC transceiver is applicability to the scenario that the phase shifts $\bm{\Theta}$ only take on finite values. At this point, the continuous phase modulated channels \eqref{eq:phase-only} are degraded to common MPSK-modulated channels, and the beamforming problem involved in \eqref{eq:siso} is converted into a discrete phase optimization problem, for which many useful algorithm are developed.
			
 \textit{4) Precoding at RIS side:}  Due to the symmetry of all RIS elements, the model \eqref{eq:equiv} can be written as
\begin{flalign}
	\check{\mathbb{Y}}&=\check{\mathsf{H}}\Pi^{-1}\cdot \Pi \exp\left(j\check{\bm{\Theta}}\right)\trans{\check{\mathbf{X}}}+\check{\mathbb{Z}}\nonumber \\
	&=\check{\mathsf{H}}^{\prime} \exp\left(j\check{\bm{\Theta}}^{\prime}\right)\trans{\check{\mathbf{X}}}+\check{\mathbb{Z}},
\end{flalign}
where $\Pi$ is an arbitrarily given permutation matrix, the permutated equivalent channel matrix $\check{\mathsf{H}}^{\prime}=\check{\mathsf{H}}\Pi^{-1}$ and the permutated phase shift vector $\check{\bm{\Theta}}^{\prime}=\Pi \check{\bm{\Theta}}$. Optimizing the permutation $\Pi$ of all RIS elements will be useful for increasing the achievable rate of the unsorted QR-SIC transceiver.

\section{Numerical Results}\label{sec:numerical}
In this section, we present numerical evaluation of our derived results. 

\subsection{Capacity Results on Hypersphere-Modulated Vector Gaussian Channels}
\begin{figure}[h]
	\centering
	\resizebox{9cm}{!}{\includegraphics{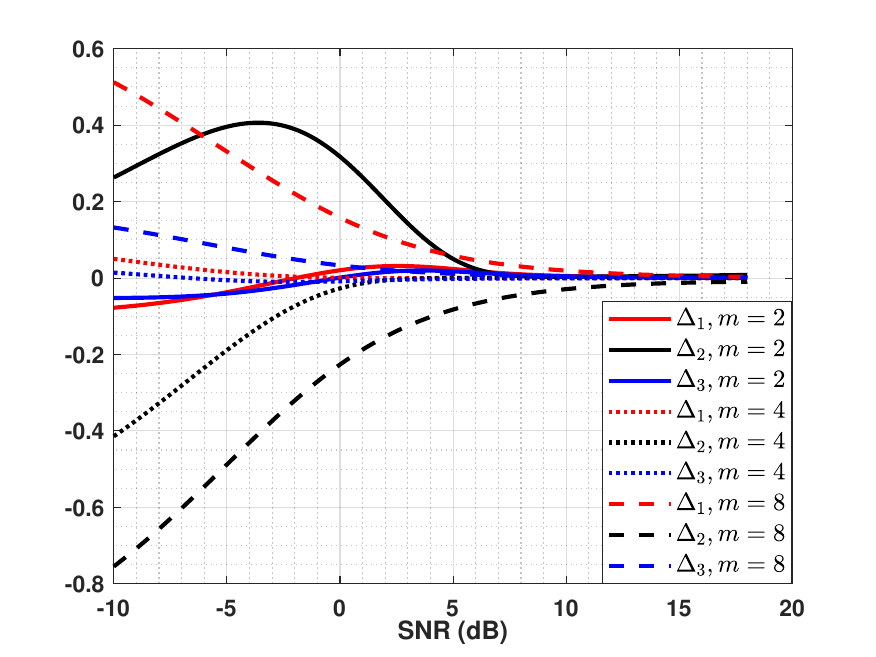}}
	\centering \caption{Scaled approximation error for $\const{K}_m(\snr)$, $\expec{\log\left( T/m \right)}$, and $\expec{R}$.}
	\label{fig:approxiamtion}
\end{figure}

We first verify our derived approximation results on the high-order expansion of functions $\const{K}_m(\snr)$ in \eqref{eq:approx_K}, the expected logarithm  $\expec{\log\left( T/m \right)}$ in \eqref{eq:approx_log}, the mean $\expec{R}$ given in \eqref{eq:appro_mean}. For notation simplicity, we let
\begin{subequations}
	\begin{align}
		\Delta_1&= \left(\const{K}_m(\snr) - \frac{-m^2+4m-3}{8m}\cdot\frac{\log e}{\snr}\right) \snr \\
			\Delta_2&=\left(\expec{\log \left(T/m\right)}-\frac{m-2}{m}\frac{\log e}{\snr}\right)\snr \\
			\Delta_3&=\left(	\expec{R} - \sqrt{m}\left(1+\frac{m-1}{2m\snr}-\frac{(m-1)(m-3)}{8m^2\snr^2}\right)\right)\snr^2,
	\end{align}
\end{subequations}
which are plotted for various dimensions $m$ in Fig.~\ref{fig:approxiamtion}. Herein, the quantities $\const{K}_m(\snr)$ and $\expec{\log \left(T/m\right)}$ are evaluated by the Monte Carlo method, and the quantity $\expec{R}$ is computed by using Eq.~\eqref{eq:mean_R}. It can be easily seen that the scaled error $\Delta_1$, $\Delta_2$, and $\Delta_3$ converges to zero as $\snr$ increases, which verifies the correctness of our derived approximation formulas \eqref{eq:approx_K}, \eqref{eq:approx_log}, and \eqref{eq:appro_mean}.

\begin{figure}[h]
	\centering
	\resizebox{9cm}{!}{\includegraphics{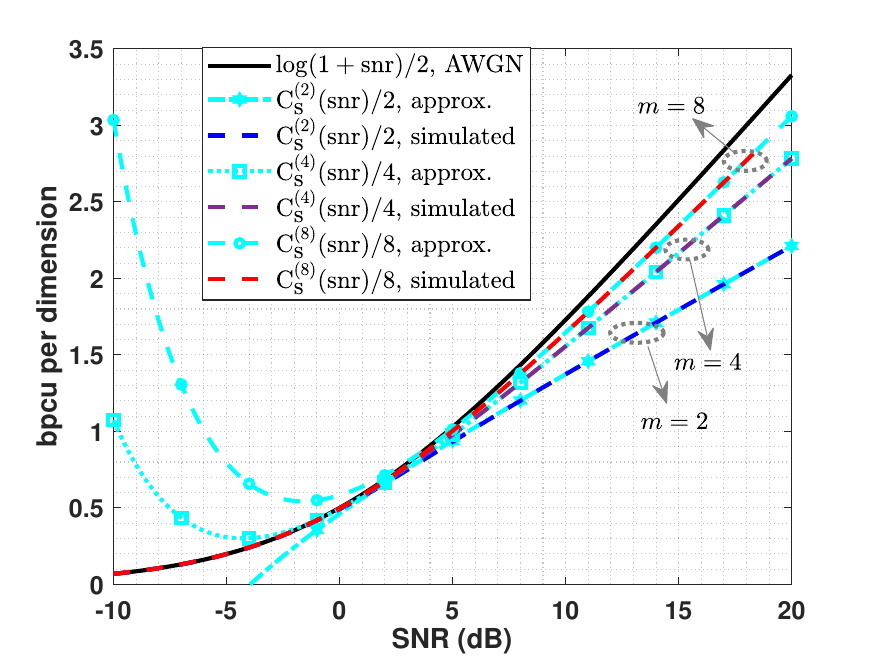}}
	\centering \caption{Capacities (per dimension) of the $2$-, $4$-, and $8$-dimensional HSM-VGCs.}
	\label{fig:HSM_VGC_capacity}
\end{figure}

Then we plot the capacities (per dimension) $\const{C}_{\textnormal{S}}^{\scriptstyle(m)}/m$ of the HSM-VGCs for various dimensions in Fig.~\ref{fig:HSM_VGC_capacity}, where the exact capacity \eqref{eq:HSMVGC_capacity} is evaluated by the Monte Carlo method and its approximation is obtained by using the high-order asymptotic formula \eqref{eq:asymptotics}. It can be seen that our derived high-SNR asymptotic capacity \eqref{eq:asymptotics} is a good match with the exact one. Moreover, an empirical observation is that we may simply but precisely evaluate $\const{C}_{\textnormal{S}}^{\scriptstyle(m)}/m$ at any SNR by the minimum one of $\frac{1}{2}\log(1+\snr)$ and the high-SNR asymptotics $\eqref{eq:asymptotics}$.  It is also noticed that, as the dimension $m$ increases, the simulated $\const{C}_{\textnormal{S}}^{\scriptstyle(m)}(\snr)/m$ approaches the capacity of the real-valued scalar AWGN channel at any SNR, which reflects the correctness of our results. 

\subsection{Capacity Results for RIS-Aided SIMO Channels}
In this subsection, we turn back to the RIS-aided SIMO AWGN channel. According to the row reduction technique introduced in Sec. \ref{subsubsec:dimension_reduction}, we assume that all involved channel matrix are defined with respect to the equivalent full-row-rank and purely-reflective model \eqref{eq:equiv}, and hence, of full-row rank. For a relatively fair comparison, all channel coefficients are independently chosen as realization of the complex Gaussian distribution $\mathcal{CN}(0,1)$ and then the obtained channel matrix $\check{\mathbb{H}}$ is normalized so that the trace of its Gram matrix $\hermi{\check{\mathbb{H}}}\check{\mathbb{H}}$ equals one. 

\begin{figure}[h]
	\centering
	\resizebox{9cm}{!}{\includegraphics{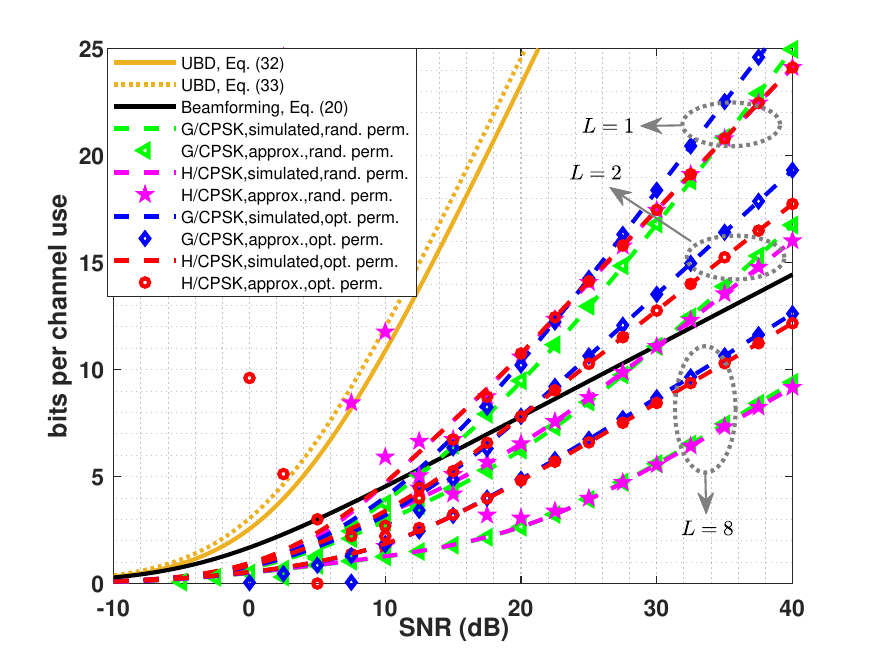}}
	\centering \caption{Capacity result for the single-input four-output channel $\check{\mathsf{H}}_1$ aided by a $4$-element RIS re-configured at different rates.}
	\label{fig:SIMO_capacity_n4tau4varyingL}
\end{figure}

In Fig. \ref{fig:SIMO_capacity_n4tau4varyingL}, for the channel matrix $\check{\mathbb{H}}_1$ (see the top of the next page) equipped with $4$ receive antennas and a $4$-element RIS with different re-configuration rates (parameterized by the SRRs $L=1$, $2$, or $8$), we plot our derived capacity upper bounds \eqref{eq:upper_bound} and \eqref{eq:upper_bound2}, maximum achievable rates of the beamforming scheme (given in Eq. \eqref{eq:AR_beamforming}) and achievable rates of our proposed QR-SIC transceiver with Gaussian/CPSK or hypersphere/CPSK input and with/without permutation optimization. It can be noticed that the proposed QR-SIC transceiver has a substantial throughput gain as compared to the beamforming scheme especially in the case of small $L$, i.e, the re-configuration rate of the RIS is near to the symbol rate of the transmitted signal $\mathbf{X}$. 
As the SRR $L$ increases, the performance advantage of the QR-SIC transceiver diminishes because of a decreased modulation rate at the RIS side.
 An interesting phenomenon is that optimizing permutation brings a considerable improvement in achievable rates of the QR-SIC transceiver expect when using hypersphere/CPSK input in the case of $L=1$. The reason for this is that, for any given non-singular square channel matrix $\check{\mathsf{H}}$ and $L=1$, the achievable rate $\mathsf{R}_{\text{QR-SIC,\,\rmnum{2}}}$ of the Hypersphere/CPSK scheme is dominated by the quantity $\const{G}_{\tau}\cdot \left(\prod_{i=1}^{\tau-1}r_i\right)=\left|\det \check{\mathsf{H}}\right|$ at high SNRs, which is permutation-invariant. In addition, there is sufficient evidence to show the achievable rates of the QR-SIC transceiver are well-approximated by our derived high-order asymptotic formulas given in Proposition \ref{prop5} and Corollary \ref{cor:2} as the SNR increases. At low SNRs, the upper bound \eqref{eq:upper_bound} coincides with the achievable rate of the beamforming scheme, which indicates the optimality of the beamforming scheme at low SNRs (as revealed by Proposition \ref{prop:low_snr_asymptotic}).
 
\begin{figure}[h]
	\centering
	\resizebox{9cm}{!}{\includegraphics{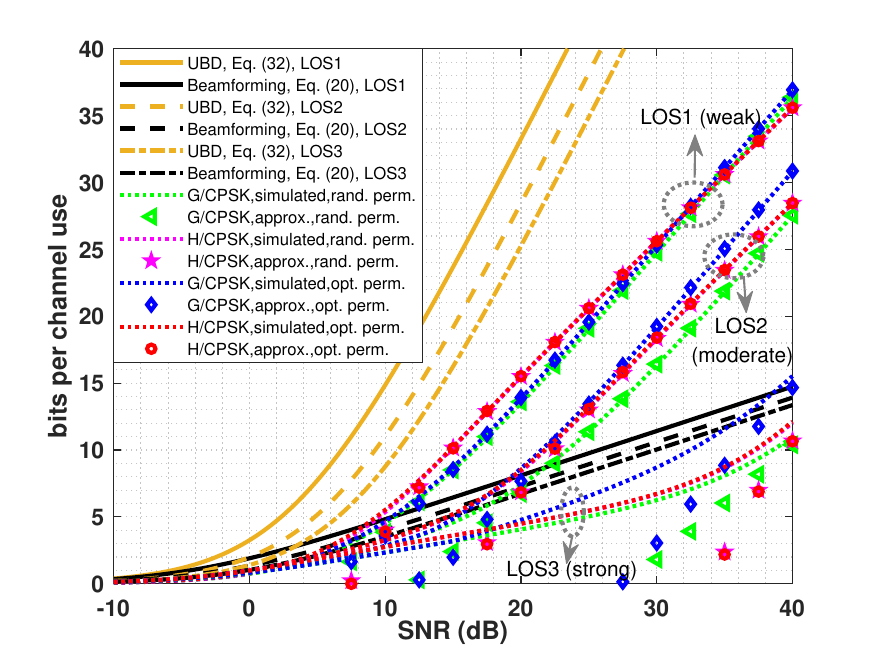}}
	\centering \caption{Capacity result for the single-input six-output channel aided by a $5$-element RIS and with a direct path under different LOS conditions.}
	\label{fig:SIMO_capacity_n6tau6L1LOS}
\end{figure}

 Next, we shall investigate the effect of direct path on the achievable rate of our concerned channel. To this end, we consider a single-input six-output channel with a $5$-element RIS and a direct path under three line-of-sight conditions, for which three channel matrices $\check{\mathsf{H}}_2$, $\check{\mathsf{H}}_3$, $\check{\mathsf{H}}_4$ are given at the top of the next page\footnote{The first channel $\check{\mathsf{H}}_2$ is under a weak LOS condition, and all entries are independently sampled from $\mathcal{CN}(0,1)$ and then normalized such that $\trace{\hermi{\check{\mathsf{H}}_2}\check{\mathsf{H}}_2}=1$. The second one is under a moderate LOS condition, and the channel matrix $\check{\mathsf{H}}_3$ is obtained by multiplying the last column vector of $\check{\mathsf{H}}_2$ by $10$ (i.e., the direct path is $10$ dB stronger than the reflected path) and then normalized to be of $\trace{\hermi{\check{\mathsf{H}}_3}\check{\mathsf{H}}_3}=1$. The last one is under a strong LOS condition, and the channel matrix $\check{\mathsf{H}}_4$ is obtained by multiplying the last column vector of $\check{\mathsf{H}}_3$ by $10$ (i.e., the direct path is $20$ dB stronger than the reflected path) and then normalized.}. Fig. \ref{fig:SIMO_capacity_n6tau6L1LOS} plots capacity upper bounds \eqref{eq:upper_bound}, maximum achievable rates of the beamforming scheme and achievable rates of the QR-SIC transceiver under above three configurations and with the SRR $L=1$. It can be seen that, as the LOS link attenuates, the performance advantage of the QR-SIC transceiver over the conventional one with optimized beamforming becomes prominent. The reason for this issue is that channels with a very strong LOS link can be well-approximated by rank-one channels, for which the beamforming scheme is optimal.

\begin{figure}[h]
	\centering
	\resizebox{9cm}{!}{\includegraphics{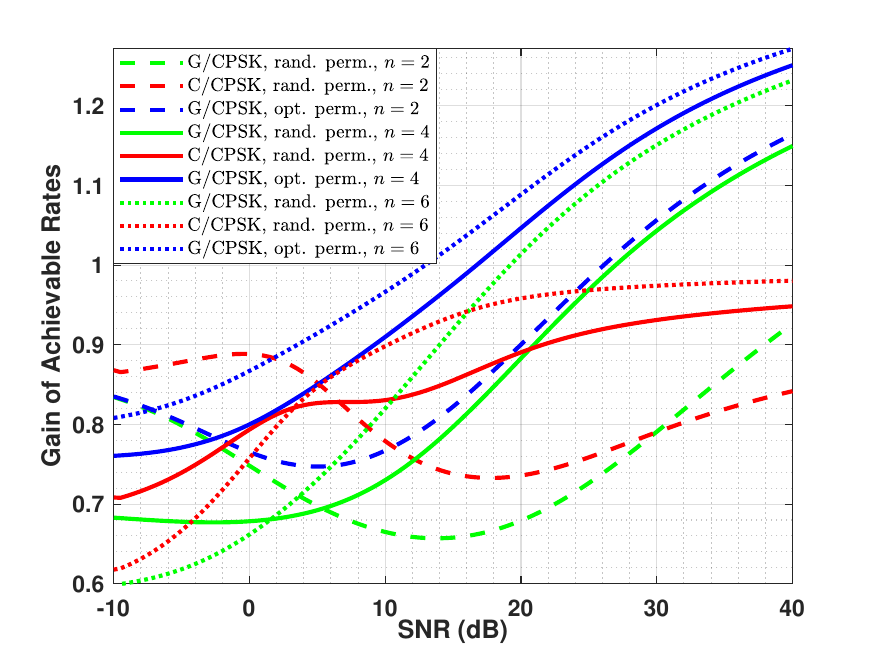}}
	\centering \caption{Average rate gain of the QR-SIC transceiver for the single-input two-output channel aided by a $n$-element RIS.}
	\label{fig:SIMO_capacity_tau2L1varyingn_fading}
\end{figure}

Finally, in Fig. \ref{fig:SIMO_capacity_tau2L1varyingn_fading} we plot average achievable rate gains (over $1000$ channel realizations) of the QR-SIC transceiver as compared to the beamforming scheme for a single-input two-output channel aided by an RIS with $n$ elements and the SRR $L=1$, From Fig. \ref{fig:SIMO_capacity_tau2L1varyingn_fading}, we observe that the rate gain increases with the number of elements, despite of the fact that the QR-SIC transceiver does not exploit all potential DoFs for spatially-correlated channel.

\newcounter{mytempeqncnt}
\begin{figure*}[!t]
	% ensure that we have normalsize text
	\normalsize
%	\caption{X}
	% Store the current equation number.
	\setcounter{mytempeqncnt}{\value{equation}}
	% Set the equation number to one less than the one
	% desired for the first equation here.
	% The value here will have to changed if equations
	% are added or removed prior to the place these
	% equations are referenced in the main text.
	\setcounter{equation}{5}
\begin{flalign*}
	\check{\mathsf{H}}_{1}=
	\begin{bmatrix}
		0.040  + 0.011i &  0.352 + 0.175i &  0.174 - 0.230i  & 0.073 + 0.158i\\
		0.074  + 0.113i & -0.392 - 0.113i & -0.289 + 0.161i & -0.290 + 0.103i\\
		-0.065  + 0.057i  & 0.387 + 0.045i & -0.102 + 0.000i &  0.082 + 0.061i\\
		-0.041  + 0.188 i &  0.059 - 0.164i & -0.048 - 0.010i & -0.211 + 0.217i
	\end{bmatrix}
\end{flalign*}
\begin{flalign*}
	\check{\mathsf{H}}_{2}=
	\begin{bmatrix}
	0.042 + 0.014i & -0.028 - 0.034i &  0.065 + 0.053i & -0.038 + 0.052i & -0.195 + 0.078i & -0.145 - 0.147i\\
	-0.074 + 0.239i & -0.224 + 0.079i & -0.195 + 0.194i &  0.166 - 0.099i  & 0.020 - 0.008i & -0.065 - 0.083i\\
	-0.041 + 0.041i  & 0.080 + 0.078i & -0.134 - 0.043i &  0.063 + 0.200i &  0.108 - 0.266i & -0.024 + 0.033i\\
	-0.075 + 0.238i & -0.016 - 0.288i & -0.059 + 0.107i &  0.155 - 0.004i & -0.039 - 0.129i &  0.006 - 0.131i\\
	-0.135 - 0.095i &  0.092 - 0.175i&   0.014 + 0.072i &  0.017 + 0.216i&  -0.071 + 0.081i&  -0.008 + 0.128i\\
	-0.120 + 0.069i &  0.036 - 0.190i &  0.149 - 0.139i & -0.087 - 0.056i & -0.041 - 0.007i &  0.081 - 0.084i
	\end{bmatrix}
\end{flalign*}
%\begin{flalign*}
%	\check{\mathsf{H}}_{3}=
%	\begin{bmatrix}
%   0.013 + 0.004i & -0.008 - 0.010i &  0.019 + 0.016i & -0.011 + 0.016i & -0.058 + 0.023i & -0.433 - 0.441i\\
%-0.022 + 0.072i & -0.067 + 0.024i  &-0.058 + 0.058i &  0.050 - 0.030i &  0.006 - 0.002i & -0.194 - 0.247i\\
%-0.012 + 0.012i &  0.024 + 0.023i  &-0.040 - 0.013i &  0.019 + 0.060i &  0.032 - 0.080i & -0.071 + 0.098i\\
%-0.022 + 0.071i & -0.005 - 0.086i  &-0.018 + 0.032i &  0.046 - 0.001i&  -0.012 - 0.039i  & 0.018 - 0.392i\\
%-0.040 - 0.029i &  0.028 - 0.052i  & 0.004 + 0.022i  & 0.005 + 0.065i & -0.021 + 0.024i&  -0.025 + 0.385i\\
%-0.036 + 0.021i  & 0.011 - 0.057i &  0.045 - 0.041i & -0.026 - 0.017i & -0.012 - 0.002i &  0.241 - 0.253i
%	\end{bmatrix}
%\end{flalign*}
%\begin{flalign*}
%	\check{\mathsf{H}}_{4}=
%	\begin{bmatrix}
%0.001 + 0.000i & -0.001 - 0.001i &  0.002 + 0.002i & -0.001 + 0.002i & -0.006 + 0.002i & -0.451 - 0.460i\\
%-0.002 + 0.007i & -0.007 + 0.002i & -0.006 + 0.006i &  0.005 - 0.003i &  0.001 + 0.000i & -0.203 - 0.258i\\
%-0.001 + 0.001i &  0.002 + 0.002i & -0.004 - 0.001i &  0.002 + 0.006i &  0.003 - 0.008i & -0.074 + 0.103i\\
%-0.002 + 0.007i &  0.000 - 0.009i & -0.002 + 0.003i &  0.005 + 0.000i  &-0.001 - 0.004i&   0.019 - 0.408i\\
%-0.004 - 0.003i &  0.003 - 0.005i &  0.000 + 0.002i  & 0.001 + 0.007i & -0.002 + 0.003i & -0.026 + 0.401i\\
%-0.004 + 0.002i &  0.001 - 0.006i &  0.005 - 0.004i  &-0.003 - 0.002i & -0.001  &  0.251 - 0.263i
%	\end{bmatrix}
%\end{flalign*}
	% Restore the current equation number.
	\setcounter{equation}{\value{mytempeqncnt}}
	% IEEE uses as a separator
	\hrulefill
	% The spacer can be tweaked to stop underfull vboxes.
	\vspace*{4pt}
\end{figure*}

\section{Conclusion}
\label{sec:conclusion}

This paper investigates the fundamental limit to the reliable communication rates achieved by using a single RF antenna and the RIS as the joint transmitter. We obtain a capacity upper bound by characterizing the maximum trace of the equivalent input. Based on this upper bound, the capacity slope at low SNRs and the exact capacity for rank-one cases are determined. It is proven that using the RIS as an optimized beamformer is optimal \rev{in the low-SNR limit or in rank-one cases}. To overcome the bottleneck of SIMO channels, we design a transceiver architecture, where the extra information is modulated on a given number of reflective elements and decoded by a QR receiver with successive interference cancellation. We characterize exact and asymptotic achievable rates of the proposed transceiver with two different signalling schemes by using our new results on the hypersphere-modulated channels. Theoretic analysis shows that the proposed transceiver indeed has an enlarged DoF than the conventional beamforming-based transceiver, of which the DoF is limited to one despite of the optimized phase shifts at the RIS side. Especially, the proposed transceiver fulfills the maximum DoF for spatially independent channels. Numerical results verify that the proposed transceiver has considerably larger achievable rates than the conventional beamforming scheme in various cases.  Above information-theoretic results somewhat provides a new perspective on the usage of the RIS for wireless communications, i.e., 1) at low SNRs, in the highly rank-deficient channels, or with a strong LOS link, the best configuration of the RIS is beamforming; 2) For channels with high SNRs and rich scattering environment, we can enlarge the throughput of an RIS-aided communication system by using (a part of) the RIS elements as phase modulators, and the simple transceiver proposed in the paper may be considered as a feasible solution.
%
% way can be well applied
%motivate us that 
%%Based on which, we characterize the asymptotic of achievable rates under two aforementioned inputs. Our proposed transceiver architecture achieves full DoF for spatially independent channels.
%Numerical results verify our derived asymptotics and show notable superiority of the proposed transceiver over the conventional-used optimized beamforming scheme.
%
%
%sideband pollution found in traditional spatiotemporally modulated metasurfaces.
%
%Systematic continuous-time signal processing. sampling.
%
%Is their exist a linear transceiver for RIS-aided communication to exploit full multiplexing gain.}

\section*{Acknowledgment} 
The authors wish to thank Prof. Longguang Li and Dr. Wenlong Guo for their valuable suggestions.

%%%%%%%%%%%%%%%%%%%%%%%%% 
%%%%%%%%%%%%%%%%%%%%%%%%%
%%%%%%%%%%%%%%%%%%%%%%%%%
\appendices
\section{Proof of Lemma \ref{lemma1}}
\label{app:proof_lemma1}
The lemma is concluded by noting that
	\begin{flalign}
		&	\II\left(\bar{\mathbf{T}}+\mathbf{Z};\bar{\mathbf{T}}\right)\nonumber \\
		=	&\hh(\bar{\mathbf{T}}+\mathbf{Z})-\hh(\mathbf{Z})  \nonumber \\
		\leq &   \log\bigl((\pi e)^{m}\det( \cov{\bar{\mathbf{T}}}+\mathsf{I}_m)\bigr)
		-\log(\pi e)^{m}  \label{eq:gaussian_maximize} \\
		=&  \log\left( \det(\mathsf{I}_m+\cov{\bar{\mathbf{T}}})\right) \\
		\leq & m\log\left(1+\frac{1}{m}\trace{\cov{\bar{\mathbf{T}}}}\right), \label{eq:ubd1}
	\end{flalign}
	where \eqref{eq:gaussian_maximize} follows from the fact that the circularly symmetric complex Gaussian distribution maximizes the differential entropy under a covariance matrix constraint, and \eqref{eq:ubd1} follows Equations (237)-(241) in \cite{limoserwangwigger20_1}. \QED

\section{Properties of Spherically Symmetric Distribution}\label{app:pre}

We first introduce the transformation from an $m$-dimensional Cartesian system to an $m$-dimensional spherical coordinate system \cite{miller1964}.
%K. S. Miller, Multidimensional Gaussian Distributions. New	York: Wiley, 1964.
Let $\mathcal{W}\triangleq[0,+\infty )\times [0,\pi]^{m-2}\times [0,2\pi)$.
Then an arbitrary real vector $\mathbf{y}\in \mathbb{R}^{m}$ can be expressed by $m$ spherical coordinates $\bm{\omega}=\left(r,\phi_1,\cdots,\phi_{m-2},\theta \right)$ through the following one-to-one transformation $\mathbb{T}^{-1}:\mathcal{W} \to \mathbb{R}^{m}$ 
\begin{flalign}
	y_i=&\, r \left(\prod_{k=1}^{m-i}\sin(\varphi_k)\right)\cos(\varphi_{m+1-i}),~i\ge 3, \nonumber \\
	y_2	=&\, r \left(\prod_{k=1}^{m-2}\sin(\varphi_k)\right)\cos(\theta), \nonumber \\
	y_1	=& \, r \left(\prod_{k=1}^{m-2}\sin(\varphi_k)\right)\sin(\theta) , 
\end{flalign}
where the radius $r\in \left[0,+\infty \right)$, angles $\varphi_1,\cdots \varphi_{m-2} \in [0,\pi]$, and the polar angle $\theta\in [0,2\pi)$. The Jacobian of $\mathbb{T}^{-1}$ is 
\begin{flalign}
	\mathsf{J}(\bm{\omega})= r^{m-1}\prod_{i=1}^{m-2}\sin^{m-1-i}\varphi_i.
\end{flalign}
For any $m$-dimensional random vector $\mathbf{Y}$, we denote its expression in a spherical coordinate system by $\bm{\Omega}=\left( R,\Phi_1,\cdots,\Phi_{m-2},\Theta \right) = \mathbb{T} \left( \mathbf{Y} \right)$. By the multivariate calculus, the joint probability density function of $\bm{\Omega}$ is given by
\begin{flalign}
	p_{\bm{\Omega}} (\bm{\omega}) =  \mathsf{J}(\bm{\omega}) \cdot p_{\mathbf{Y}}\left(\mathbb{T}^{-1}\left( \bm{\omega} \right)\right).
\end{flalign}

Then we turn to a specific class of continuous distributions whose probability density at any point $\mathbf{y}$ is only determined by its $\ell_2$-norm $\left\|\mathbf{y} \right\|_2$, termed as \textit{spherically symmetric distribution} (or \textit{rotationally invariant distribution}).  The following proposition presents good properties of spherically symmetric distributions.
\begin{proposition}[\cite{goldman1976detection,karout2017ofc}]
	Let $\bm{\Omega}=\left( R,\Phi_1,\cdots,\Phi_{m-2},\Theta \right)$ be the expression of an $m$-dimensional spherically symmetric random vector $\mathbf{Y}$ in the spherical coordinate system. Then
	\begin{enumerate}
		\item[\textbf{P1:}] $R,\Phi_1,\cdots,\Phi_{m-2},\Theta$ are mutually independent;
		\item[\textbf{P2:}] For $i\in [m-2]$, the angle $\Phi_i$ is distributed as 
		\begin{flalign}
			p_{\Phi_i}(\phi_i)=\frac{\Gamma(\frac{m-i+1}{2})}{\Gamma(\frac{m-i}{2})\sqrt{\pi}}\sin^{m-i-1}(\phi_i);
		\end{flalign}
		\item[\textbf{P3:}] The polar angle $\Theta$ is uniformly distributed, i.e.,
		\begin{flalign}
			p_{\Theta}(\theta)=\frac{1}{2\pi};
		\end{flalign}
		\item[\textbf{P4:}] The marginal distribution of $R$ satisfies
		\begin{flalign}\label{eq:P4}
			p_{\scriptscriptstyle \! R}\left( \left\| \mathbf{y}  \right\|_2 \right) 	 = \const{A}_{m}\left\| \mathbf{y}  \right\|_2^{m-1}\cdot p_{\mathbf{Y}} (\mathbf{y}),
		\end{flalign} 	
		where the constant $\const{A}_{m}=\frac{2\pi^{m/2}}{\Gamma(m/2)}$ is the surface area of the unit $m$-sphere $\mathcal{S}_{m-1}$ in Euclidean $m$-space.
	\end{enumerate}
\end{proposition}

\section{Proof of Theorem \ref{thm:2}} \label{app:proof_thm2}

The proof is completed by the following two steps.

\textit{Step 1: Optimality of the uniform distribution.} Let the random orthogonal matrix $\mathbb{Q}\in \mathbb{R}^{m\times m}$ is equiprobably chosen from the orthogonal group $\mathcal{O}_m$ (with respect to the Haar measure). For any channel input $\mathbf{X}$ and any realization $\mathbb{Q}=\mathsf{Q}$, it can be easily verified that the rotated input $\tilde{\mathbf{X}}= \mathsf{Q}\mathbf{X}$ satisfies the desired support constraint $\mathsf{supp} \, \tilde{\mathbf{X}} \subseteq \sqrt{m}\mathcal{S}_{m-1}$. By the rotational invariance of the standard Gaussian vector, we have $\II(\tilde{\mathbf{X}};\sqrt{\mathsf{snr}}\cdot\tilde{\mathbf{X}}+\mathbf{Z})=\II({\mathbf{X}};\sqrt{\mathsf{snr}}\cdot{\mathbf{X}}+\mathbf{Z})$. Let $\mathbf{U} \triangleq \mathbb{Q}\mathbf{X}$, then conditional uniformly-distributed property immediately shows that $\mathbf{U}$ is uniformly distributed on $m$-sphere $\sqrt{m}\mathcal{S}_{m-1}$. Then the optimality of the uniform distribution $\mathbf{U}$ can be shown by
\begin{flalign}
	&\II(\mathbf{U};\sqrt{\mathsf{snr}}\cdot\mathbf{U}+\mathbf{Z})\nonumber \\
	=& \II	(\mathbb{Q}\mathbf{X};\sqrt{\mathsf{snr}}\cdot\mathbb{Q}\mathbf{X}+\mathbf{Z} \nonumber \\
	\ge& \, \mathbb{E}_{\mathbb{Q}}\left[ \II(\mathbb{Q}\mathbf{X};\sqrt{\mathsf{snr}}\cdot\mathbb{Q}\mathbf{X}+\mathbf{Z})\vert \mathbb{Q} \right] \label{eq:concavity} \\
	=&  \II(\mathbf{X};\sqrt{\mathsf{snr}}\cdot\mathbf{X}+\mathbf{Z}),
\end{flalign}
where Eq. \eqref{eq:concavity} follows from the concavity of mutual information functional.

\textit{Step 2: Calculate $\II(\mathbf{U};\sqrt{\mathsf{snr}}\cdot\mathbf{U}+\mathbf{Z})$.}
For facilitation of asymptotic expansion, we equivalently consider the scaled channel output $\mathbf{Y}^\prime= \mathbf{U}+\mathbf{Z}/\sqrt{\mathsf{snr}}$.

Denote the $\ell_2$-norm and the squared $\ell_2$-norm of $\mathbf{Y}^\prime$ by $R$ and $T$, respectively. For any $\mathbf{u}\in \sqrt{m}\mathcal{S}_{m-1}$, there exists orthogonal matrix $\mathsf{Q}$ such that $\bm{1}_m = \mathsf{Q}\mathbf{u}$. Combining with the spherical symmetry of standard Gaussian random vector, we immediately know that $T$ is a non-central chi-squared distribution $\chi^2_m(m,\snr)$ with the probability density function given in \eqref{eq:pdf_chisquared}, and $R$ is a generalized Ricean random variable \cite{Proakis2008}.

%
%For any finite number $k$, a well-known convergent () Hadamard expansion of the Bessel function is given by
%\begin{flalign}
%	I_v(x)&= \frac{e^x}{\sqrt{2 \pi x}}\sum_{k=0}^{+\infty}\frac{a_k(v)}{(2x)^k}P(\frac{1}{2}+v+k,2x)\nonumber \\
%	&\sim  \frac{e^x}{\sqrt{2 \pi x}}\sum_{k=0}^{+\infty}\frac{a_k(v)}{(2x)^k} \nonumber \\
%	&\sim \frac{e^x}{\sqrt{2 \pi x}}
%\end{flalign} 

It can be easily verified that the sum of any two independent spherically symmetric random vector is also spherically symmetric. Hence, the channel output $\mathbf{Y}^\prime$ is symmetrically symmetric, and its differential entropy can be simplified by using the method in \cite{karout2017ofc} as follows
\begin{flalign}
	&\hh(\mathbf{Y}^\prime)  \nonumber\\
	=&-\int_{\mathbb{R}^{m}}
	p_{ \mathbf{Y}^\prime}(\mathbf{y})\log p_{\mathbf{Y}^\prime}(\mathbf{y}) \dd \mathbf{y} \nonumber \\
	=&-\int_{\mathbb{R}^{m}}
	\frac{p_{\scriptscriptstyle \! R}\left( \left\| \mathbf{y}  \right\|_2 \right)}{\const{A}_{m}\left\| \mathbf{y}  \right\|_2^{m-1}}
	\log \frac{p_{\scriptscriptstyle \! R}\left( \left\| \mathbf{y}  \right\|_2 \right)}{\const{A}_{m}\left\| \mathbf{y}  \right\|_2^{m-1}} \dd \mathbf{y} \nonumber \\
	=&-\int_{0}^{+\infty} \dd r \int_{r\mathcal{S}_{  m \! - \! 1}} 
	\frac{p_{\scriptscriptstyle \! R}\left( r \right)}{\const{A}_{m}r^{m-1}}
	\log \frac{p_{\scriptscriptstyle \! R}\left(r \right)}{\const{A}_{m}r^{m-1}} \dd S    \label{eq:joint2marginal}\\
	=&-\int_{0}^{+\infty} 
	p_{\scriptscriptstyle \! R}\left( r \right)
	\log \frac{p_{\scriptscriptstyle \! R}\left(r \right)}{\const{A}_{m}r^{m-1}} \dd r \\
	=& \hh(R)+\log(\const{A}_m)+(m-1)\mathbb{E}\left[ \log R \right] \nonumber\\
	=&-1+ \hh(T)+\log(\const{A}_m)+\frac{m-2}{2}\mathbb{E}\left[ \log T \right] \label{eq:transform_entropy}
\end{flalign}
where $p_{\scriptscriptstyle \! R}(r)$ is the probability density function of $R$, $\dd S$ denotes the infinitesimal of $m$-sphere $r\mathcal{S}_{m-1}$, Eq.~\eqref{eq:joint2marginal} follows from Eq.~\eqref{eq:P4}, and Eq. \eqref{eq:transform_entropy} follows from using variable transformation in the integral involved in the differential entropy.

Using \eqref{eq:pdf_chisquared}, the differential entropy of the chi-squared variable $T$ can be expressed as
\begin{flalign}
	\hh(T)=&1-\log \snr -\frac{m-2}{4}\expec{\log \frac{T}{m}}+\frac{ \log e}{2}\left(\expec{T}+m\right)\snr\nonumber \\
	&-\expec{\log I_{\frac{m}{2}-1}(\sqrt{mT}\snr)} \label{eq:entropy_T}
\end{flalign}
Combining Eqs. \eqref{eq:transform_entropy} and \eqref{eq:entropy_T}, we obtain
\begin{flalign}
	&\const{C}_{\textrm{S}}^{(m)}(\mathsf{snr})\nonumber\\
	=&\II(\mathbf{U};\mathbf{Y}^\prime) \nonumber \\
	=&\hh(\mathbf{Y}^\prime)-\hh(\mathbf{Z}/\sqrt{\snr}) \nonumber\\
	=&-\log \snr +\frac{m-2}{4}\expec{\log  T}+ \frac{m-2}{4}\log m \nonumber \\
	&+\left(m\snr +\frac{m}{2}\right)\log(e)-\expec{\log I_{\frac{m}{2}-1}(\sqrt{mT}\snr)} \nonumber\\
	&+\log \const{A}_m  + \frac{m}{2}\log\left(\frac{\snr}{2\pi e} \right)\\
=	&\left(\frac{m}{2}-1\right)\log\frac{\snr}{2 }  +\frac{m-2}{4}\expec{\log mT}+\left(m\snr \right)\log(e) \nonumber \\
	&-\expec{\log I_{\frac{m}{2}-1}(\sqrt{mT}\snr)} -\log(\Gamma(m/2)).
\end{flalign}
This is the complete proof of Theorem \ref{thm:2}. \QED

\section{Proof of Theorem \ref{thm:3}} \label{app:proof_thm3}
We first review that $I_v(x)$ denotes the $v$-th order modified Bessel function of the first kind, whose definition and a well-know absolutely convergent Hadamard expansion for the case of $\rp{v}>-1/2$  are given as follows:
\begin{flalign}
	I_v(x)&\triangleq \sum_{k=0}^{+\infty} \frac{1}{k!\Gamma(v+k+1)}\left(\frac{x}{2}\right)^{v+2k} \nonumber \\
	&= \frac{e^x}{\sqrt{2 \pi x}}\sum_{k=0}^{+\infty}\frac{a_k(v)}{(2x)^k}P\left(\frac{1}{2}+v+k,2x\right) \label{eq:expansion_Bessel}\\
	&\sim  \frac{e^x}{\sqrt{2 \pi x}}\sum_{k=0}^{+\infty}\frac{a_k(v)}{(2x)^k} ,
\end{flalign}
where $P(\mu,x)$ denotes the normalized incomplete gamma function and converges to unity as $x\to \infty$ for finite $\mu$, coefficients $a_k(v)$ are given by
\begin{flalign}
	a_k(v)=\frac{\left(\frac{1}{2}+v\right)_k\left(\frac{1}{2}-v\right)_k}{k!}
\end{flalign}
and $(\alpha)_k\triangleq\alpha(\alpha+1)\cdots(\alpha+k-1)$ denotes the Pochhammer symbol \cite{paris2004jcam}.

Then we define the function $\const{K}_m(\snr)$ as
\begin{flalign}
&\const{K}_m(\snr) \nonumber \\
\triangleq & \,\expec{\log I_{\frac{m}{2}-1}(\sqrt{mT}\snr)} -\expec{\log\frac{\exp(\sqrt{mT}\snr)}{\sqrt{2 \pi \sqrt{mT}\snr}}}.
\end{flalign}
The channel capacity \eqref{eq:HSMVGC_capacity} can be represented as 
\begin{flalign}
	&\const{C}_{\textrm{S}}^{(m)}(\mathsf{snr})\nonumber\\
	=	&\left(\frac{m}{2}-1\right)\log\frac{\snr}{2 }  +\frac{m-2}{4}\expec{\log mT}+\left(m\snr \right)\log(e) \nonumber \\
	&-\const{K}_m(\snr)-\log(e)\sqrt{m}\snr\expec{\sqrt{T}}+\frac{1}{2}\log\left( 2\pi \snr\right)\nonumber\\
	&+\frac{1}{4}\expec{\log mT} -\log(\Gamma(m/2))\\
		=	&\frac{m-1}{2}\log\frac{\snr}{2 }  +\frac{m-1}{2}\log m+\log\left( \frac{2\sqrt{\pi}}{\Gamma(m/2)} \right) \nonumber \\
	&-\log(e)\sqrt{m} \left(\expec{R}-\sqrt{m}\right)\snr+\frac{m-1}{4}\expec{\log \frac{T}{m}}\nonumber\\
	&  -\const{K}_m(\snr) .\label{eq:95}
\end{flalign}

Note that the following series
\begin{flalign}
	\ln(1+x)=	\sum_{k=1}^{+\infty}(-1)^{k-1}\frac{1}{k}x^k.
\end{flalign}
is uniformly convergent when $|x|<1$. Then, by using Hadamard expansion \eqref{eq:expansion_Bessel} of the modified function $I_v(\cdot)$, we immediately obtain the asymptotics of $\const{K}_m(\snr)$ as follows\footnote{Note that here we directly expand the series of $\log(1+x)$. A more strict (but tedious) approach is truncating $T$ in a closed interval around $m$ and combining its exponential decay property.   }
\begin{flalign}
	&\const{K}_m(\snr) \nonumber \\
	=\,&\expec{\log\left( 1+ \sum_{k=1}^{+\infty}\frac{a_k(\frac{m}{2}-1)}{(2\sqrt{mT}\snr)^k}P\left(\frac{m-1}{2}+k,2\sqrt{mT}\snr\right)\right)} \nonumber \\
	=\,& \expec{\frac{a_1(m/2-1)\log(e)}{2\sqrt{mT}\snr}}+o\left(\frac{1}{\snr}\right) \nonumber \\
	=\, & \frac{-m^2+4m-3}{8m}\cdot\frac{\log e}{\snr}+o\left(\frac{1}{\snr}\right). \label{eq:approx_K}
\end{flalign}
The expected logarithm of the non-central chi-squared varibale $T$ satisfies\footnote{See \cite{moser2020chisquare} for a closed-form expression.}
\begin{flalign}
	&\expec{\log \left(T/m\right)}\nonumber \\
	=&\expec{\log \left( 1+\frac{\left\| \mathbf{Z} \right\|_2^2}{m \snr}+\frac{2}{m\sqrt{\snr}}\left( \sum_{i=1}^m Z_i\right) \right)} \nonumber \\
	=& \frac{m-2}{m}\frac{\log e}{\snr} + o\left(\frac{1}{\snr}\right).\label{eq:approx_log}
\end{flalign}
According to \cite[Eq. (2.3-66)]{Proakis2008}, the mean of the generalized Ricean random variable $R$ is given by
\begin{flalign}
	&\expec{R}\nonumber \\
	=&\sqrt{\frac{2}{\snr}}\exp(-\frac{m\snr}{2})\frac{\Gamma(\frac{m+1}{2})}{\Gamma(\frac{m}{2})}\cdot     {}_{1}F_1\left(\frac{m+1}{2},\frac{m}{2};\frac{m \snr}{2}\right),\label{eq:mean_R}
%	\\
%	=& \sqrt{m}\left( \sum_{k=0}^{+\infty} \frac{\left(\frac{1-m}{2}\right)_{k}(-\frac{1}{2})_k}{k! \left( m\snr/2 \right)^k}P\left(k-\frac{1}{2},\frac{m \snr}{2}\right) \right) 
\end{flalign}
where ${}_1F_1(\cdot,\cdot;\cdot)$ denotes the confluent hypergeometric function. Note that the Hadamard expansion for ${}_1F_1(\cdot,\cdot;\cdot)$ takes the following form \cite{paris2004jcam}
\begin{flalign}
&\frac{\Gamma(a)}{\Gamma(b)} \cdot  {}_1F_1(a,b;x) \nonumber \\
=&	x^{a-b}e^x \left(\sum_{k=0}^{+\infty} \frac{(1-a)_k(b-a)_k}{k!x^k} P(b-a+k,x) \right) ,
\end{flalign}
and hence, we have
\begin{flalign}
	&\expec{R}-\sqrt{m}\nonumber \\
	=& \sqrt{m}\left(\frac{m-1}{2m\snr}-\frac{(m-1)(m-3)}{8m^2\snr^2}+o\left(\frac{1}{\snr^2   }\right)\right) .\label{eq:appro_mean}
\end{flalign}

Substituting Eqs. \eqref{eq:approx_K}, \eqref{eq:approx_log}, and \eqref{eq:appro_mean} into Eq.~\eqref{eq:95}, then the high-SNR asymptotic expression of the channel capacity $\const{C}_{\textrm{S}}^{(m)}(\mathsf{snr})$ is immediately obtained as in Eq.~\eqref{eq:asymptotics}.
\QED

\section{Proof of Proposition \ref{prop5}}\label{app:prop5} 	
In \cite{moser2020chisquare}, the expected logarithm and negative-order moments of a central chi-squared variable are given by
\begin{flalign}\label{eq:expected_logarithm1}
	\expec{\ln \hat{T}} = \ln 2 +\psi \left( L \right),
\end{flalign}
and
\begin{flalign}\label{eq:negative_moment}
	\expec{ \hat{T}^{-1}} =\begin{cases}
		\infty&,~L=1,\\
		\frac{1}{2(L-1)}&,~L\ge1,
	\end{cases}
\end{flalign}
respectively, where $\psi(\cdot)$ denotes the digamma function 
\begin{flalign}\label{eq:digamma}
	\psi \left( L \right) = -\gamma + \sum_{j=1}^{L-1}\frac{1}{j}
\end{flalign}
and $\gamma$ denotes the Euler constant. Then we note that the CDF function $F_{\hat{T}}(t)$ of the central chi-squared variable $\hat{T}$ increases at a order of $ t^{L-1}$ as $t\to 0^+ $. For above two reasons, there is some difference in infinitesimal analysis between the case of $L=1$ and that of $L\ge2$.

%\begin{flalign}
%	&	\expec{\const{C}_{\textrm{S}}^{(2)}\left(\frac{r_{ii}^2 \mathcal{E} \hat{T}}{2} \right)}  \nonumber\\
%	=& \expec{\frac{1}{2}\log\left(\frac{2\pi}{e} r_{ii}^2 \mathcal{E} \hat{T}\right) -\frac{\log(e)}{4r_{ii}^2 \mathcal{E} }
	%	}+o\left(\frac{1}{\mathcal{E}}\right)
%\end{flalign}

In the case of $L=1$, by using Eqs.~\eqref{eq:wyner_asymptotic},~\eqref{eq:expected_logarithm1}, and~\eqref{eq:digamma}, we have
\begin{flalign}
	&	\expec{\const{C}_{\textrm{S}}^{(2)}\left(\frac{r_{ii}^2 \mathcal{E} \hat{T}}{2} \right)}  \nonumber\\
	=& \expec{\frac{1}{2}\log\left(\frac{2\pi}{e} r_{ii}^2 \mathcal{E} \hat{T}\right)  
	}+o\left(1\right)  \nonumber \\
	=&\frac{1}{2}\log\left(\frac{2\pi}{e} r_{ii}^2 \mathcal{E} \right)+\frac{1}{2}\expec{\log\hat{T}}+o(1) \nonumber \\
	=&\frac{1}{2}\log\left(\frac{2\pi}{e} r_{ii}^2 \mathcal{E} \right)+\frac{1+\log(e) \psi(1)}{2} +o(1) \label{eq:72}\\
	=&\frac{1}{2}\log\left(\frac{4\pi}{e} r_{ii}^2 \mathcal{E} \right)-\frac{ \log(e) \gamma}{2} +o(1). \label{eq:73}
\end{flalign}

In the case of $L\ge2 $, by using Eqs.~\eqref{eq:our_asymptotic},~\eqref{eq:expected_logarithm1}, \eqref{eq:negative_moment}, and~\eqref{eq:digamma}, we have
\begin{flalign}\label{eq:74}
	&	\expec{\const{C}_{\textrm{S}}^{(2)}\left(\frac{r_{ii}^2 \mathcal{E} \hat{T}}{2} \right)}  \nonumber\\
	=&\expec{\frac{1}{2}\log\left(\frac{2\pi}{e} r_{ii}^2 \mathcal{E} \hat{T}\right) -\frac{\log(e)}{4r_{ii}^2 \mathcal{E} \hat{T} }}+o\left(\frac{1}{\mathcal{E}}\right)\nonumber \\
	=&\frac{1}{2}\log\left(\frac{2\pi}{e} r_{ii}^2 \mathcal{E} \right)+\frac{1}{2}\expec{\log\hat{T}}-\frac{\log(e)}{4r_{ii}^2 \mathcal{E} }\expec{\hat{T}^{-1}}+o\left( \frac{1}{\mathcal{E}} \right)\nonumber \\
	=&\frac{1}{2}\log\left(\frac{4\pi}{e} r_{ii}^2 \mathcal{E} \right)+\frac{ \log(e) \psi(L) }{2}-\frac{\log(e)}{8\left(L-1\right)r_{ii}^2 \mathcal{E} }+o\left( \frac{1}{\mathcal{E}} \right).
\end{flalign}
The proposition is concluded by substituting Eqs. \eqref{eq:73} and \eqref{eq:74} into Eq. \eqref{eq:rate_1}. \QED

\bibliographystyle{ieeetr}
\bibliography{./defshort1,./biblio1}

%%%%%%%%%%%%%%%%%%%%%%%%%%%%%%%%%%%%%%%%%%%%%%%%%%%%%%%%%%%%%%%%%%%%%
\addtolength{\textheight}{-68mm}

\end{document}